\documentclass[preprint, a4paper, 10pt, oneside, onecolumn]{elsarticle}
\usepackage[left=2.5cm, right=2.5cm, top=2.5cm, bottom=2.5cm]{geometry}

\makeatletter
\def\ps@pprintTitle{%
 \let\@oddhead\@empty
 \let\@evenhead\@empty
 \def\@oddfoot{Manuscript to appear in the {\em Journal of Mathematical Psychology} \hfill}
 \let\@evenfoot\@oddfoot}
\makeatother

\usepackage{lineno}

\usepackage{multicol}

\usepackage{natbib}
\setcitestyle{authoryear}

\usepackage{algorithm2e}
\usepackage{xcolor}

\usepackage{amsmath,amssymb,mathtools}
\usepackage{bbm}

\newcommand{\EE}{\mathbb{E}}

\newcommand{\PP}{\mathbb{P}}
\newcommand{\RR}{\mathbb{R}}

\newcommand{\Nn}{\mathcal{N}}
\newcommand{\Pp}{\mathcal{P}}
\newcommand{\Ss}{\mathcal{S}}

\newcommand{\Xx}{\mathcal{X}}
\newcommand{\Yy}{\mathcal{Y}}

\newcommand{\DKL}{D_{\rm KL}}

\newcommand{\flatp}{{\rm flat}}

\newcommand{\tS}[1]{_{#1}}
\newcommand{\tSb}[1]{^{(#1)}}

\usepackage{amsthm}

\newtheorem{mydef}{Definition}
\newtheorem{mylemma}{Lemma}
\newtheorem{myprop}{Proposition}
\newtheorem{myremark}{Remark}

\newtheorem{mycorrol}{Corollary}

\usepackage[colorlinks]{hyperref}

\definecolor{mygreen}{RGB}{0, 120, 0}
\definecolor{myred}{RGB}{120, 0, 0}
\definecolor{myblue}{RGB}{0, 0, 120}

\hypersetup{
  citecolor=red,
  linkcolor=red,
  urlcolor=red}

\usepackage{pifont}

\usepackage{verbatim}

\newcommand{\parencite}[1]{\citep{#1}}

\begin{document}
\begin{frontmatter}
\title{A taxonomy of surprise definitions\tnoteref{t1}}
\tnotetext[t1]{See \cite{modirshanechi2021surp} for experimental predictions.}
\author[1]{Alireza Modirshanechi\corref{cor1}}
\author[1]{Johanni Brea}
\author[1]{Wulfram Gerstner}
\cortext[cor1]{Corresponding author: alireza.modirshanechi@epfl.ch}
\address[1]{EPFL, School of Computer and Communication Sciences and School of Life Sciences, Lausanne, Switzerland}

\begin{abstract}
    Surprising events trigger measurable brain activity and influence human behavior by affecting learning, memory, and decision-making. Currently there is, however, no consensus on the definition of surprise. Here we identify 18 mathematical definitions of surprise in a unifying framework. We first propose a technical classification of these definitions into three groups based on their dependence on an agent’s belief, show how they relate to each other, and prove under what conditions they are indistinguishable. Going beyond this technical analysis, we propose a taxonomy of surprise definitions and classify them into four conceptual categories based on the quantity they measure: (i) `prediction surprise' measures a mismatch between a prediction and an observation; (ii) `change-point detection surprise' measures the probability of a change in the environment; (iii) `confidence-corrected surprise' explicitly accounts for the effect of confidence; and (iv) `information gain surprise' measures the belief-update upon a new observation. The taxonomy poses the foundation for principled studies of the functional roles and physiological signatures of surprise in the brain.
\end{abstract}


\end{frontmatter}

\tableofcontents

\section{Introduction}
Imagine you open the curtains one morning and find the street in front of your apartment covered by fresh snow.
If you have expected a warm and sunny morning according to the weather forecast, you feel `surprised' as you see the white streets;
as a consequence of surprise, the activity of many neurons in your brain changes \parencite{squires1976effect, mars2008trial, kolossa2015computational} and your pupils dilate \parencite{antony2021behavioral, preuschoff2011pupil, nassar2012rational}.
Surprise affects how we predict and perceive our future and how we remember our past.
For example, some studies suggest that you would rely less on the weather forecast for your future plans after the snowy morning \parencite{behrens2007learning, nassar2010approximately, xu2021novelty}.
Other studies predict that you would remember more vividly the face of the random stranger who walked past the street in that very moment you felt surprised \parencite{rouhani2018dissociable, rouhani2021signed}, and some predict that this moment of surprise might have even modified your memory of another snowy morning in the past \parencite{gershman2017computational, sinclair2018surprise}.
To understand and explain the computational role of surprise in different brain functions, one first needs to ask `what does it really mean to be surprised?' and formalize how surprise is perceived by our brain.
For instance, when you see the white street, do you feel `surprised' because what you expected turned out to be wrong \parencite{meyniel2016human, faraji2018balancing, glascher2010states} or because you need to change your trust in the weather forecast \parencite{baldi2002computational, schmidhuber2010formal, liakoni2019approximate}?

Computational models of perception, learning, memory, and decision-making often assume that humans implicitly perceive their sensory observations as probabilistic outcomes of a generative model with hidden variables
\parencite{angela2005uncertainty, friston2010free, fiser2010statistically, gershman2017computational, soltani2019adaptive, findling2019imprecise,  liakoni2019approximate}.
In the example above, the observation is whether it snows or not and the hidden variables characterise how the probability of snowing depends on old observations and relevant context information (such as the current season, yesterday’s weather, and the weather forecast).
Different brain functions are then modeled as aspects of statistical inference and probabilistic control in such generative models \parencite{angela2005uncertainty, behrens2007learning, glascher2010states, daw2011model, nassar2012rational,  gershman2017computational, meyniel2016human, friston2017active, findling2019imprecise,  dubey2019reconciling, liakoni2019approximate, horvath2020human}.
In these probabilistic settings, surprise of an observation depends on the relation between the observation and our expectation of what to observe.

In the past decades, different definitions and formal measures of surprise have been proposed and studied \parencite{baldi2002computational, glascher2010states, schmidhuber2010formal, friston2010free, palm2012novelty, barto2013novelty, kolossa2015computational, faraji2018balancing, liakoni2019approximate}.
These surprise measures have been successful both in explaining the role of surprise in different brain functions \parencite{itti2006bayesian, gershman2017computational, xu2021novelty, rouhani2021signed, antony2021behavioral, findling2019imprecise} and in identifying signatures of surprise in behavioral and physiological measurements \parencite{mars2008trial, glascher2010states, rubin2016representation, modirshanechi2019trial, maheu2019brain, gijsen2021}.
However, there are still many open questions including, but not limited to:
(i) Are the quantities that different definitions of surprise measure conceptually different?
(ii) Can we identify mathematical relations between different surprise definitions?
In particular, is one definition a special case of another one, completely distinct, or do they have some common ground?

In this work, we analyze and discuss 18 previously proposed measures of surprise in a unifying framework.
We first present our framework, assumptions, and notation in \autoref{sec:gen_model}.
Then, in \autoref{sec:surprise_theory} to \autoref{sec:surprise_theory_belief}, we give definitions for each of the 18 surprise measures and show their similarities and differences.
In particular, we identify conditions that make different surprise measures experimentally indistinguishable.
Finally, in \autoref{sec:conc_class}, we build upon our theoretical analyses and propose a taxonomy of surprise measures by classifying them into four conceptually different categories.

\section{Subjective world-model: A unifying generative model}
\label{sec:gen_model}
\begin{figure*}[!t]
    \centering
    \includegraphics[width=1\textwidth]{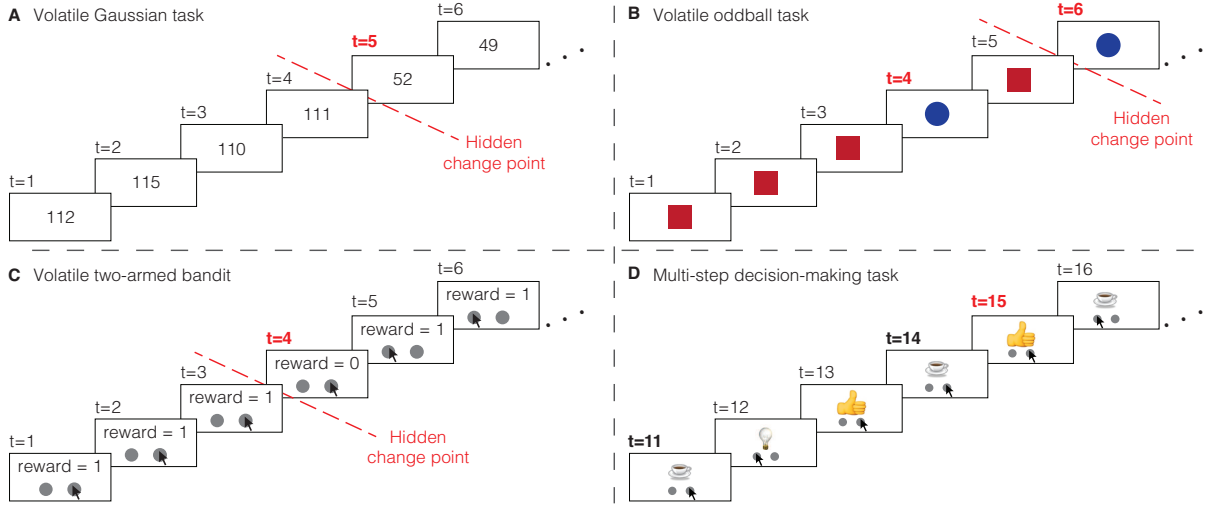}
    \caption{\textbf{Four typical experimental paradigms to study functional roles and physiological signatures of surprise in the brain.}
    \textbf{A.} Volatile Gaussian task \parencite{nassar2010approximately, nassar2012rational}:
    Participants see a sequence of numbers randomly sampled from a Gaussian distribution whose mean is piece-wise constant but abruptly changes at random points in time (change-points, e.g., $t = 5$ in the figure).
    The goal of participants is to predict the next observation; hence, the first few observations after a change-point are unexpected.
    Variants of this paradigm have been studied by \cite{o2013dissociable} and \cite{visalli2021EEG}.
    \textbf{B.} Volatile oddball task \parencite{heilbron2019confidence, meyniel2020brain}:
    Participants see a sequence of binary stimuli (e.g., a red square and a blue disk).
    The stimulus frequencies are piece-wise constant but abruptly change at random points in time (change-points, e.g., $t = 6$ in the figure).
    During the stationary periods between two consecutive change-points (before $t=6$ in the figure), one stimulus (the blue disk, called `deviant') is less frequent than the other (the red square, called `standard') and hence more surprising than the other.
    Variants of the paradigm with more than 2 types of stimuli \parencite{mars2008trial,lieder2013modelling} or without change-points \parencite{huettel2002perceiving, maheu2019brain, modirshanechi2019trial, squires1976effect}  have also been studied.
    \textbf{C.} Volatile two-armed bandit task \parencite{behrens2007learning, horvath2020human}:
    Participants select one action (e.g., click on one of the grey disks in the figure) at a time and receive a reward value randomly sampled from a distribution specific to the selected action.
    The reward distributions are piece-wise stationary but switch at random change points (e.g., $t = 4$ in the figure).
    Participants optimize reward and have to adapt their strategy after a change-point.
    Variants of the paradigm include, e.g., multi-dimensional actions \parencite{niv2015reinforcement} or context-dependent reward distributions \parencite{rouhani2021signed}.
    \textbf{D.} Multi-step decision-making task \parencite{glascher2010states, xu2021novelty, liakoni2022brain}:
    Participants move between states (e.g., images of different objects) by selecting one action (e.g., clicking on one of the disks in the figure) at a time.
Assuming some transitions have been experienced before (e.g., the `light bulb' state followed by selecting the right action in the `cup' state), observing the `light bulb' state at $t = 12$ is expected, whereas observing the `thumb' state at $t = 15$ after the same stimulus-action sequence at $t = 14$ as at $t = 11$ is unexpected and hence surprising.
    \textit{Color should be used in print.}
    }
    \label{fig:experiments}
\end{figure*}

Our goal is to study the theoretical properties of different formal measures of surprise in a common mathematical framework.
To do so, we need to make assumptions on how an agent (e.g., a human participant or an animal) thinks about its environment.
We assume that an agent thinks of its observations as probabilistic outcomes of a generative model with hidden variables and, hence, consider a generative model that captures several key features of daily life and unifies many existing model environments in neuroscience and psychology (c.f. \autoref{sec:app_special_cases}).
More specifically, we assume that the generative model describes the subjective interpretation of the environment from the point of view of the agent and, importantly, that the agent takes the possibility into account that the environment may undergo abrupt changes at unknown points in time (i.e., the environment is volatile), similar to the experimental paradigms studied by \cite{behrens2007learning, nassar2010approximately, glaze2015normative, heilbron2019confidence, xu2021novelty, maheu2019brain}. See \autoref{fig:experiments} for four typical experimental paradigms that are used to study behavioral and physiological signatures of surprise.
Note that we do not assume that the environment has the same dynamics as those assumed by the agent.

\subsection{General definition}

\begin{figure*}[!t]
    \centering
    \includegraphics[width=1\textwidth]{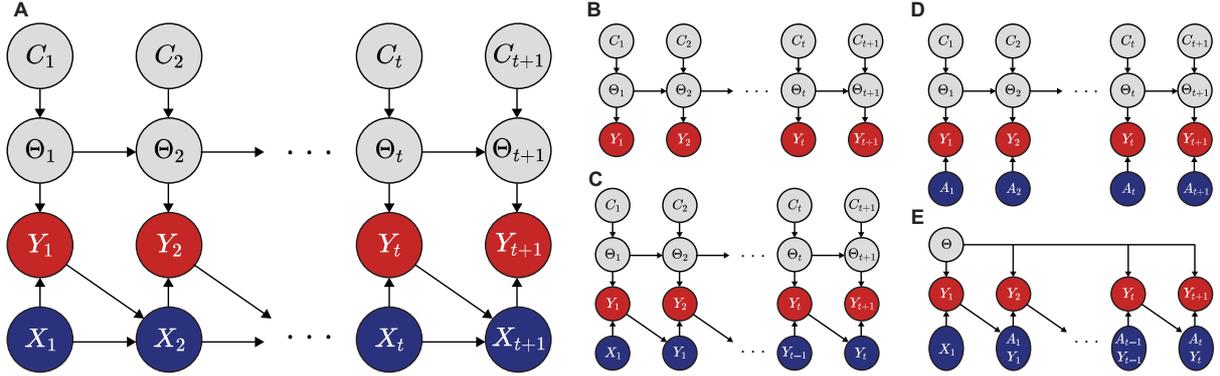}
    \caption{\textbf{Subjective model of the environment.}
    \textbf{A.} The Bayesian network \parencite{barber2012bayesian} corresponding to the most general case of our generative model in \autoref{eq:gen_mod_1} and \autoref{eq:gen_mod_2}.
    The arrows show conditional dependence, the grey nodes show the hidden variables ($C\tS{1:t+1}$ and $\Theta\tS{1:t+1}$), the red nodes show the observations ($Y\tS{1:t+1}$), and the blue nodes show the cue variables ($X\tS{1:t+1}$).
    A variety of tasks can be written in the form of a reduced version of our generative model.
    Specifically:
    \textbf{B.} Standard generative model for modeling and studying passive learning in experiments with volatile environments like the one in \autoref{fig:experiments}A \parencite{adams2007bayesian, fearnhead2007line, nassar2010approximately, nassar2012rational, wilson2013mixture, liakoni2019approximate},
    \textbf{C.} generative model for modeling human inference about binary sequences in experiments like the one in \autoref{fig:experiments}B \parencite{meyniel2016human, maheu2019brain,modirshanechi2019trial, mousavi2020brain,gijsen2021},
    \textbf{D.} generative model corresponding to variants of bandit and volatile bandit tasks like the one in \autoref{fig:experiments}C \parencite{behrens2007learning, findling2019imprecise, horvath2020human}, where the cue variable $X\tS{t} = A\tS{t}$ is a participant's action,
    and \textbf{E.} classic Markov Decision Processes (MDPs) to model experiments like the one in \autoref{fig:experiments}D \parencite{sutton2018reinforcement,schultz1997neural, glascher2010states, daw2011model, huys2015interplay, lehmann2019one}, where the cue variable $X\tS{t} = (A\tS{t-1}, Y\tS{t-1})$ consists of previous action and observation.
    See \autoref{sec:app_special_cases} for details.
    \textit{Color should be used in print.}
    }
    \label{fig:gen_model}
\end{figure*}

At each discrete time $t \in \{0,1,2,...\}$, the agent's model of the environment is characterized by a tuple of 4 random variables $(X\tS{t},Y\tS{t},\Theta\tS{t},C\tS{t})$ (\autoref{fig:gen_model}A).
$X\tS{t}$ and $Y\tS{t}$ are observable, whereas $\Theta\tS{t}$ and $C\tS{t}$ are unobservable (hidden).
We refer to $X\tS{t}$ as the cue and to $Y\tS{t}$ as the observation at time $t$.
Examples of an observation are an image on a computer screen \parencite{mars2008trial, kolossa2015computational} (e.g., \autoref{fig:experiments}), an auditory tone \parencite{imada1993determinants, lieder2013modelling}, and an electrical stimulation \parencite{ostwald2012evidence}.
The cue variable $X\tS{t}$ can be interpreted as a predictor of the next observation, since it summarizes the necessary information needed for predicting the observation $Y\tS{t}$.
Examples of a cue variable are the previous observation $Y\tS{t-1}$ \parencite{meyniel2016human, modirshanechi2019trial}, the last action of a participant (which we will denote by $A\tS{t-1}$) \parencite{behrens2007learning, horvath2020human} (e.g., \autoref{fig:experiments}C-D), and a conditioned stimulus in Pavlovian conditioning tasks \parencite{gershman2017computational}.

At time $t$, given the cue variable $X\tS{t}$, the agent assumes that the observation $Y\tS{t}$ comes from a distribution that is conditioned on $X\tS{t}$ and is parameterized by the hidden variable $\Theta\tS{t}$.
We do not put any constraints on the sets to which $X\tS{t}$, $Y\tS{t}$, and $\Theta\tS{t}$ belong.
We refer to $\Theta\tS{t}$ as the environment parameter at time $t$.
The sequence of variables $\Theta\tS{1:t} = (\Theta_1, ...,\Theta_t)$ describe the temporal dynamics of the observations $Y\tS{1:t}$ given the cue variables $X\tS{1:t}$ in the agent's model of the environment.
Similar to well-known models of volatile environments \parencite{angela2005uncertainty, angela2009sequential, behrens2007learning, adams2007bayesian, fearnhead2007line, nassar2010approximately, nassar2012rational, wilson2013mixture, glaze2015normative, meyniel2016human, heilbron2019confidence, findling2019imprecise, liakoni2019approximate, xu2021novelty}, the agent assumes that the environment undergoes abrupt changes at random points in time (e.g., \autoref{fig:experiments}A-C).
An abrupt change at time $t$ is specified by the event $C\tS{t}=1$ and happens with a probability $p_c \in [0,1)$; otherwise $C\tS{t}=0$.
If the environment abruptly changes at time $t$ (i.e., $C\tS{t}=1$), then the agent assumes that the environment parameter $\Theta\tS{t}$ is sampled from a prior distribution $\pi\tSb{0}$ independently of $\Theta\tS{t-1}$;
if there is no change ($C\tS{t}=0$), then $\Theta\tS{t}$ remains the same as $\Theta\tS{t-1}$.
We refer to $p_c$ as the change-point probability.

We use $\PP$ to refer to probability distributions:
Given a random variable $W$ and a value $w \in \RR$, we use $\PP(W=w)$ to refer to the probability of event $\{W=w\}$ for discrete random variables and, with a slight abuse of notation, to the probability density function of $W$ at $W=w$ for continuous random variables.
In general, we denote random variables by capital letters and their values by small letters.
However, for any pair of arbitrary random variables $W$ and $V$ and their values $w$ and $v$, whenever there is no risk of ambiguity, we either drop the capital- or the small-letter notation and, for example, write $\PP(W=w|V=v)$ as $\PP(w|v)$.
When there is a risk of ambiguity, we keep the capital notation for the random variables, e.g., we write $\PP(W=v,V=v)$ as $\PP(W=v,v)$.
Given this convention, the agent's model of the environment described above is formalized in Definition \ref{def:gen_model} (c.f. \autoref{fig:gen_model}A).

\begin{mydef}
\label{def:gen_model}
(Subjective world-model)
An agent's model of the environment is defined for $t>0$ as a joint probability distribution over $Y\tS{1:t}$, $X\tS{1:t}$, $\Theta\tS{1:t}$, and $C\tS{1:t}$ as
\begin{linenomath*} \begin{equation}\ \begin{aligned}
\label{eq:gen_mod_1}
\PP &\big(y\tS{1:t},x\tS{1:t},\theta\tS{1:t},c\tS{1:t}\big) \coloneqq\,
\PP\big(c\tS{1}\big)
\PP\big(\theta\tS{1}\big)
\PP\big(x\tS{1}\big)
\PP\big(y\tS{1}|x\tS{1},\theta\tS{1}\big) \times \\
&\prod_{\tau = 2}^t
\PP\big(c\tS{\tau}\big)
\PP\big(\theta\tS{\tau}|\theta\tS{\tau-1},c\tS{\tau}\big) \PP\big(x\tS{\tau}|x\tS{\tau-1},y\tS{\tau-1}\big)
\PP\big(y\tS{\tau}|x\tS{\tau},\theta\tS{\tau}\big),
\end{aligned} \end{equation} \end{linenomath*}
where $c\tS{1}$ is by definition equal to $1$ (i.e., $\PP(c\tS{1}) \coloneqq \delta_{\{1\}}(c\tS{1})$),
$\PP\big(\theta_1\big) \coloneqq \pi\tSb{0}(\theta_1)$ for an arbitrary distribution $\pi\tSb{0}$, and
\begin{linenomath*} \begin{equation}\ \begin{aligned}
\label{eq:gen_mod_2}
\PP(c\tS{\tau}) \coloneqq &{\rm Bernoulli}(c\tS{\tau}; p_c)\\
\PP\big(\theta\tS{\tau}|\theta\tS{\tau-1},c\tS{\tau}\big) \coloneqq & \pi\tSb{0}(\theta\tS{\tau}) \delta_{\{1\}}(c\tS{\tau}) + \delta_{\{\theta\tS{\tau-1}\}}(\theta\tS{\tau}) \delta_{\{0\}}(c\tS{\tau})\\
\PP\big(y\tS{\tau}|x\tS{\tau},\theta\tS{\tau} \big) \coloneqq &P_{Y|X}(y\tS{\tau}|x\tS{\tau};\theta\tS{\tau}),
\end{aligned} \end{equation} \end{linenomath*}
where $\delta$ is the Dirac measure (c.f. \autoref{tab:notation}), and $P_{Y|X}$ is a time-invariant conditional distribution of observations given cues\footnote{The last line of \autoref{eq:gen_mod_2} implies that $\PP\big(Y\tS{\tau} = y|X\tS{\tau} = x,\Theta\tS{\tau} = \theta \big) = \PP\big(Y\tS{\tau'} = y|X\tS{\tau'} = x,\Theta\tS{\tau'} = \theta \big) = P_{Y|X}(y|x;\theta)$ for any $\tau$ and $\tau' \in \{0,1,2,...\}$.}.
We do not make any assumption about $\PP\big(x\tS{1}\big)$ and $\PP\big(x\tS{\tau}|x\tS{\tau-1},y\tS{\tau-1}\big)$.\end{mydef}

See \autoref{tab:notation} for a summary of the notation.

\begin{table*}[!t]
\centering
\caption{\label{tab:notation} Notation summary}
\begin{tabular}{l||l}
Notation & Meaning \\\hline\hline
$X\tS{t}$ & Cue at time $t$ \\\hline
$Y\tS{t}$ & Observation at time $t$ \\\hline
$\Theta\tS{t}$ & Environment parameter at time $t$ \\\hline
$C\tS{t}$ & Change-point indicator at time $t$ \\\hline
$p_c$ & Change-point probability, i.e., the probability of $C\tS{t}=1$ \\\hline
$P_{Y|X}(y|x;\theta)$ & Time invariant distribution of observation $y$ given cue $x$, parameterized by $\theta$\\\hline
$\PP$ & The distribution corresponding to the subjective model of the environment; see Definition \ref{def:gen_model}  \\\hline
$\PP\tSb{t}$ & $\PP$ conditioned on observations and cues until time $t$, i.e., $x\tS{1:t}$ and $y\tS{1:t}$ \\\hline
$\PP\tSb{t}_W$ & An alternative notation for the distribution of random variable $W$\\
               & conditioned on $x\tS{1:t}$ and $y\tS{1:t}$, i.e., $\PP\tSb{t}_{W}(w) \coloneqq \PP\tSb{t}(W = w)$\\\hline
$\pi\tSb{0}$ & Prior distribution over the environment parameter; equivalently, the distribution of $\Theta\tS{t}$\\
             & given $C\tS{t} = 1$ \\\hline
$\pi\tSb{t}$ & The belief about parameter $\Theta\tS{t}$ at time $t$, i.e., $\pi\tSb{t}(\theta) \coloneqq \PP\tSb{t}(\Theta\tS{t} = \theta)$ \\\hline
$P(y|x;\pi\tSb{t})$ & The marginal probability of observation $y$ given cue $x$ and belief $\pi\tSb{t}$; see \autoref{eq:marginal_prob}\\\hline
$P(.|x;\pi\tSb{t})$ & The full marginal distribution over the space of observations given cue $x$ and belief $\pi\tSb{t}$\\\hline
$||w||_1$ & $\ell_1$-norm of the vector $w = (w_1,...,w_N) \in \RR^N$ defined as $||w||_1 \coloneqq \sum_{n=1}^N |w_n|$\\\hline
$||w||_2$ & $\ell_2$-norm of the vector $w = (w_1,...,w_N) \in \RR^N$ defined as $||w||_2 \coloneqq \sqrt{\sum_{n=1}^N w_n^2}$\\\hline
$\delta_{\{w^*\}}$ & The Dirac measure at $w^*$, i.e., $\PP(W=w) = \delta_{\{w^*\}}(w)$ implies that the probability of the \\
                   & event $\{W=w^*\}$ is one.
\end{tabular}
\end{table*}

\subsection{Special cases and links to related works}
\label{sec:app_special_cases}
Many of the commonly used experimental paradigms (e.g., see \autoref{fig:experiments}) can be formally described in our framework as special cases of Definition \ref{def:gen_model}.
The standard generative models for studying passive learning in volatile environments \parencite{adams2007bayesian, nassar2010approximately, nassar2012rational, liakoni2019approximate} is obtained if we remove the cue variables $X\tS{1:t}$ (\autoref{fig:gen_model}B).
For example, in the Gaussian experiment of \cite{nassar2010approximately} (\autoref{fig:experiments}A), $Y\tS{t}$ is a sample from a Gaussian distribution with a mean equal to $\Theta\tS{t}$ and a known variance, and $\pi\tSb{0}$ is a very broad uniform distribution.

The minimal model of human inference about binary sequences of \cite{meyniel2016human} (\autoref{fig:gen_model}C) assumes that participants estimate probabilities of transitions between stimuli instead of stimulus frequencies, even when the stimuli are by design independent of each other.
They show that such an assumption helps explaining many experimental phenomena.
Their model is obtained as a special case of our generative model if the cue variable $X\tS{t}$ is equal to the previous observation $Y\tS{t-1}$.
There, $Y\tS{t}$, conditioned on $Y\tS{t-1}$, is a sample from a Bernoulli distribution with parameter $\Theta\tS{t}$.
In this setting, we have $\PP\big(x\tS{\tau}|x\tS{\tau-1}, \allowbreak y\tS{\tau-1}\big) \coloneqq \delta_{\{y\tS{\tau-1}\}}( x\tS{\tau})$.
This class of generative models has been used to study the neural signatures of surprise via encoding \parencite{maheu2019brain, gijsen2021} and decoding \parencite{modirshanechi2019trial} models in oddball tasks (\autoref{fig:experiments}B).

Variants of bandit and reversal bandit tasks \parencite{behrens2007learning, findling2019imprecise, horvath2020human} can be modeled by considering the cue variables $X\tS{1:t}$ as actions $A\tS{1:t}$ (\autoref{fig:gen_model}D).
For example, in the experiment of \cite{behrens2007learning} (\autoref{fig:experiments}C), $X\tS{t}=A\tS{t}$ is one of the two possible actions that participants can choose, $Y\tS{t}$ is the indicator of whether they are rewarded or not, and $\Theta\tS{t}$ indicates which action is rewarded with higher probability.
In this setting, $\PP\big(x\tS{\tau}|x\tS{\tau-1},y\tS{\tau-1}\big) = \PP\big(x\tS{\tau}\big)$ is the probability that participants take action $x\tS{\tau}$, independently of the dynamics of the environment\footnote{We note that the action probability $\PP\big(a\tS{\tau}\big)$ in bandit tasks often depends on the whole history of the agent, i.e., $a\tS{1:\tau-1}$ and $y\tS{1:\tau-1}$ \parencite{sutton2018reinforcement}.
In these situations, one can define $x\tS{\tau}$ as the concatenation of $a\tS{1:\tau}$ and $y\tS{1:\tau-1}$.
In this case, the dynamics are described by
$\PP\big(X\tS{\tau} = (a'\tS{1:\tau}, y'\tS{1:\tau-1})|x\tS{\tau-1},y\tS{\tau-1}\big) \coloneqq
\delta_{\{a\tS{1:\tau-1}\}}(a'\tS{1:\tau-1}) \allowbreak
\delta_{\{y\tS{1:\tau-1}\}}(y'\tS{1:\tau-1})
\PP\big(a'\tS{\tau} | a\tS{1:\tau-1}, y\tS{1:\tau-1} \big)$ where $\PP\big( a'\tS{\tau}  | \allowbreak a\tS{1:\tau-1}, y\tS{1:\tau-1} \big)$ is the non-stationary action selection policy -- c.f. \cite{sutton2018reinforcement}.}.

Classic Markov Decision Processes (MDPs) \parencite{sutton2018reinforcement} can also be written in the form of our generative model.
To reduce our generative model to an MDP, we set $p_c=0$, consider the observation $Y\tS{t}$ as the pair of the current state and immediate reward value, and consider the cue variable $X\tS{t}$ as the previous pair of action and observation (or state) $(A\tS{t-1}, Y\tS{t-1})$ (\autoref{fig:gen_model}E).
In this setting, we have $\PP\big(X\tS{\tau}=(a\tS{\tau-1},y)|x\tS{\tau-1},y\tS{\tau-1}\big) \coloneqq \allowbreak \delta_{\{ y\tS{\tau-1}\}}(y) \allowbreak  \PP\big(a\tS{\tau-1}|y\tS{\tau-1}\big)$, where $\PP\big(a\tS{\tau-1}|y\tS{\tau-1}\big)$ is called the action selection policy in Reinforcement Learning theory \parencite{sutton2018reinforcement} and is independent of the dynamics of the environment\footnote{Similar to the case of bandit tasks, action selection policies in reinforcement learning algorithms used for solving MDPs often depend on the sequence of previous actions $a\tS{1:\tau-1}$ and observations $y\tS{1:\tau-1}$, i.e., through estimation of action values \parencite{sutton2018reinforcement}.
In these situations, we can define $x\tS{\tau}$ as the concatenation of $a\tS{1:\tau}$ and $y\tS{1:\tau-1}$.}.
The theory of Reinforcement Learning for MDPs has been frequently used in neuroscience and psychology to model human reward-driven decision-making \parencite{glascher2010states, daw2011model, huys2015interplay, niv2009reinforcement, lehmann2019one, xu2021novelty} (\autoref{fig:experiments}D).

\subsection{Additional notation, belief, and marginal probability}

We define $\PP\tSb{t}$ as $\PP$ conditioned on the sequences of observations $y\tS{1:t}$ and cue variables $x\tS{1:t}$.
For example, for an arbitrary random variable $W$ with value $w$, we write $\PP\tSb{t}(w) \coloneqq \PP(w|y\tS{1:t},x\tS{1:t})$.
Following this notation, we define an agent's belief about the parameter $\Theta_t$ at time $t$ as
\begin{linenomath*} \begin{equation}\ \begin{aligned}
\label{eq:belief}
\pi\tSb{t}(\theta) \coloneqq \PP\tSb{t}(\Theta_t = \theta),
\end{aligned} \end{equation} \end{linenomath*}
that is the posterior probability (or density, for continuous $\Theta_t$) of $\Theta_t = \theta$ conditioned on $y\tS{1:t}$ and $x\tS{1:t}$.
The belief plays a crucial role in the perception of surprise (c.f. \autoref{sec:tech_class}), and we assume that an agent constantly updates its belief, through either exact or approximate Bayesian inference, as it makes new observations --
see \cite{barber2012bayesian} and \cite{liakoni2019approximate} for examples of inference algorithms in generative models similar to ours.
According to exact Bayesian inference \parencite{barber2012bayesian}, the updated belief $\pi\tSb{t+1}(\theta) = \PP\tSb{t+1}(\Theta\tS{t+1} = \theta)$ can be found by normalizing the product of the \textit{prior} belief $\PP\tSb{t}(\Theta\tS{t+1} = \theta)$ about $\Theta_{t+1}$ and the \textit{likelihood} $P_{Y|X}(y\tS{t+1}|x\tS{t+1};\theta)$.
In \autoref{sec:surprise_theory_prob_BF}, we give a simple and interpretable expression of the updated belief for the generative model of Definition \ref{def:gen_model} (c.f. Proposition \ref{Prop:update}).

Another important quantity is the marginal probability of observing $y$ given the cue $x$ and a belief $\pi\tSb{t}$:
\begin{linenomath*} \begin{equation}\ \begin{aligned}
\label{eq:marginal_prob}
P(y|x;\pi\tSb{t}) &\coloneqq \EE_{\pi\tSb{t}} \Big[ P_{Y|X}(y|x;\Theta) \Big]\\
&= \int P_{Y|X}(y|x;\theta) \pi\tSb{t}(\theta) d \theta,
\end{aligned} \end{equation} \end{linenomath*}
where the integration is replaced by summation whenever $\theta$ is discrete.

\section{Surprise measures and indistinguishability}
\label{sec:surprise_theory}
\begin{figure*}[t]
    \centering
    \includegraphics[width=1\textwidth]{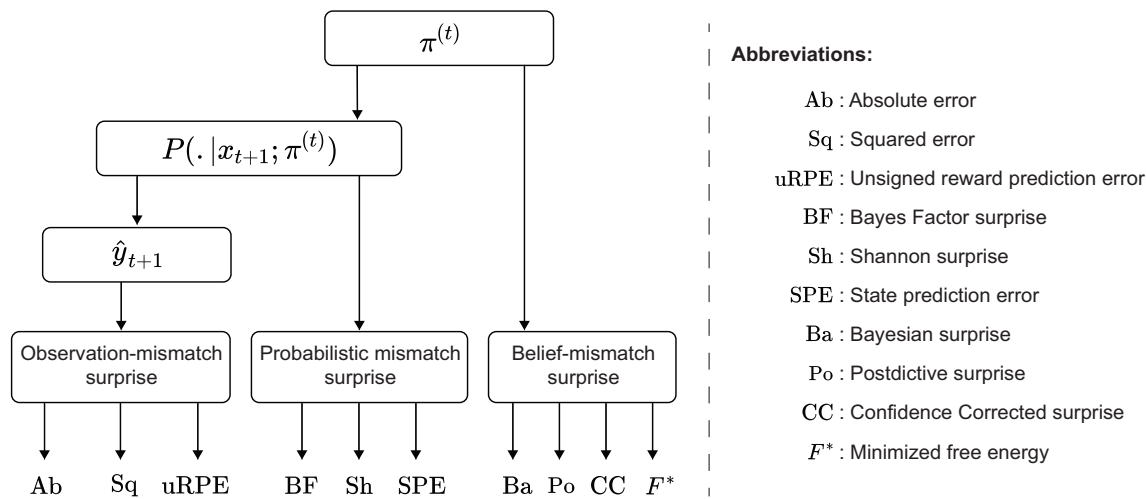}
    \caption{\textbf{Technical classification of surprise measures based on the form of their dependence upon the agent's belief.}
    Surprise depends on expectations.
    Therefore, all surprise measures depend on the belief $\pi\tSb{t}$.
    However, the specific form of the dependence changes between one measure and another.
    `Observation-mismatch' surprise measures use the marginal distribution $P(.|x\tS{t+1}; \pi\tSb{t})$ (c.f. \autoref{tab:notation}) to calculate an estimate $\hat{y}\tS{t+1}$ of the next observation, which is then compared with the real observation $y\tS{t+1}$ by an error function such as $||\hat{y}\tS{t+1} - y\tS{t+1}||_1$ (c.f. \autoref{tab:notation}).
    `Probabilistic mismatch' surprise measures use the marginal probability $P(y\tS{t+1}|x\tS{t+1}; \pi\tSb{t})$ directly, without extracting a specific estimate.
    `Belief-mismatch' surprise measures use the belief $\pi\tSb{t}$ directly, without extracting the marginal probability $P(y\tS{t+1}|x\tS{t+1}; \pi\tSb{t})$.
See \autoref{sec:surprise_theory} for details.}
    \label{fig:surprise_belief}
\end{figure*}

Conditioned on the previous observations $y\tS{1:t}$ and cue variables $x\tS{1:t+1}$, how surprising is the next observation $y\tS{t+1}$?
We address this question by examining previously proposed measures of surprise.
In this section, we propose a technical classification of different surprise measures and a notion of indistinguishability between different measures and, in the next three sections, we define all surprise measures in the same mathematical framework and discuss their differences and similarities.
We present the proofs of these results in \ref{sec:app_proofs}.

\subsection{A technical classification}
\label{sec:tech_class}

Given $\theta\tS{t+1}$, the observation $y\tS{t+1}$ is independent of the previous observations $y\tS{1:t}$ and cue variables $x\tS{1:t}$ and only depends on $x\tS{t+1}$ (\autoref{fig:gen_model}A).
Hence, the influence of $y\tS{1:t}$ and $x\tS{1:t}$ on the surprise of observing $y\tS{t+1}$ is exclusively through the belief $\pi\tSb{t}$, which indicates the importance of $\pi\tSb{t}$ in surprise computation.
More precisely, a surprise measure is a function $\Ss: \Yy \times \Xx \times \Pp \to \RR$ that takes an observation $y\tS{t+1} \in \Yy$, a cue $x\tS{t+1} \in \Xx$, and a belief $\pi\tSb{t} \in \Pp$ as arguments and gives the value $\Ss(y\tS{t+1}|x\tS{t+1}; \pi\tSb{t}) \in \RR$ as the corresponding surprise value.
However, the specific form of how $\pi\tSb{t}$ influences surprise computation changes between one measure and another.
Based on how they depend on $\pi\tSb{t}$, we divide existing surprise measures into three categories: (i) probabilistic mismatch, (ii) observation-mismatch, and (iii) belief-mismatch surprise measures (\autoref{fig:surprise_belief}).
\textit{Probabilistic mismatch} surprise measures depend on the belief $\pi\tSb{t}$ only through the marginal probability $P(y\tS{t+1}|x\tS{t+1}; \pi\tSb{t})$;
an example is the Shannon surprise \parencite{barto2013novelty, tribus1961thermostatics}.
In other words, probabilistic mismatch surprise depends only on the integral  $P(y\tS{t+1}| \allowbreak x\tS{t+1};\pi\tSb{t}) = \int P_{Y|X}(y\tS{t+1}|x\tS{t+1};\theta) \pi\tSb{t}(\theta) d \theta$ (\autoref{eq:marginal_prob}) and is independent of other characteristics of the belief $\pi\tSb{t}$.
\textit{Observation-mismatch} surprise measures depend on $\pi\tSb{t}$ only through some estimate $\hat{y}\tS{t+1}$ of the next observation according to the marginal distribution $P(.|x\tS{t+1};\pi\tSb{t})$ (c.f. \autoref{tab:notation});
an example is the absolute difference between $y\tS{t+1}$ and $\hat{y}\tS{t+1}$ \parencite{nassar2010approximately, prat2021human}.
In other words, observation-mismatch surprise depends only on some statistics (e.g., average or mode) of $P(.|x\tS{t+1};\pi\tSb{t})$ that is used as the estimate $\hat{y}\tS{t+1}$ and is independent of the other characteristics of $\pi\tSb{t}$ and $P(.|x\tS{t+1};\pi\tSb{t})$.
To compute the \textit{belief-mismatch} surprise measures, however, we need to have the whole distribution $\pi\tSb{t}$;
an example is the Bayesian surprise \parencite{baldi2002computational, schmidhuber2010formal}.
In other words, neither the marginal distribution $P(.|x\tS{t+1};\pi\tSb{t})$ nor the estimate $\hat{y}\tS{t+1}$ can solely determine the value of a belief-mismatch surprise measure.

\subsection{Notion of indistinguishability}

\begin{figure*} [t!]
  \centering
  \includegraphics[width=1\textwidth]{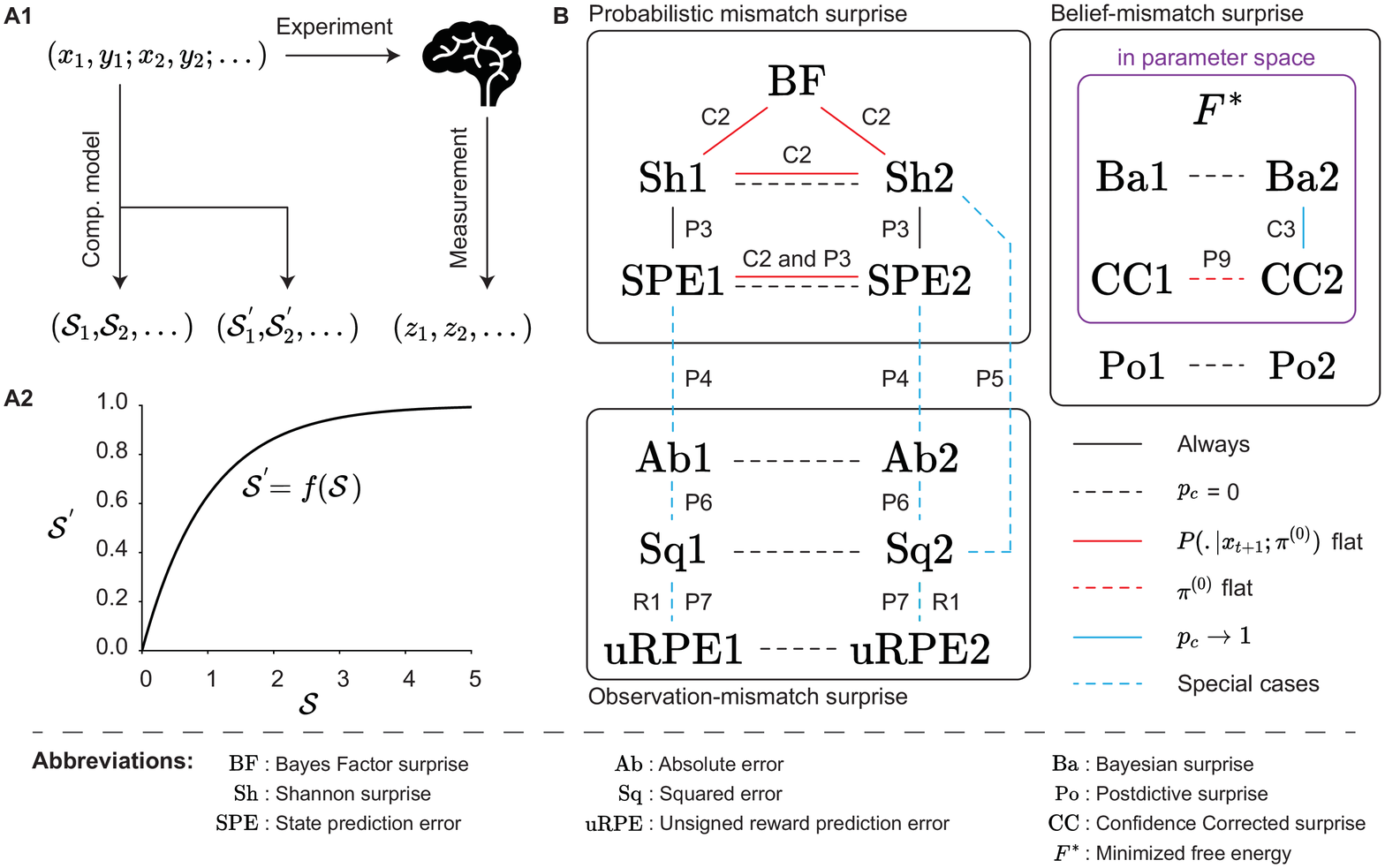}
  \caption{\textbf{Indistinguishable surprise measures.}
    \textbf{A.} A typical question in human and animal experiments is whether a surprise measure $\Ss$ explains the variations of a behavioral or physiological variable $Z$ better than an alternative surprise measure $\Ss'$.
    \textbf{A1.} A common experimental paradigm:
    A sequence of cues $x\tS{1:t}$ and observations $y\tS{1:t}$ is presented to participants, the sequence $z\tS{1:t}$ is measured, and the sequence of surprise values $\Ss\tS{1:t}$ or $\Ss'\tS{1:t}$ is predicted by computational modeling.
Then statistical tools are used to study whether the sequence $\Ss\tS{1:t}$ or $\Ss'\tS{1:t}$ is more informative about the sequence of measurements $z\tS{1:t}$.
    \textbf{A2.} If there exists a strictly increasing function $f$ such that $\Ss' = f(\Ss)$, then the two surprise measures are equally informative about the measurable variable $Z$.
    In this case, $\Ss$ and $\Ss'$ are `indistinguishable' (c.f. Definition \ref{def:indis}).
    \textbf{B.} Schematic of the theoretical relation between different measures of surprise.
    A line connecting two measures indicates that the two measures are indistinguishable, i.e., one is a strictly increasing function of the other, under the condition corresponding to the color and the type of the line.
    The conditions are shown on the bottom right of the panel:
    a solid black line means the two measures are always indistinguishable;
    a dashed black line corresponds to the condition $p_c=0$;
    a solid red line corresponds to the prior marginal probability $P(.|x\tS{t+1}; \pi\tSb{0})$ being flat;
    a dashed red line corresponds to the prior belief  $\pi\tSb{0}$ being flat;
a solid blue line corresponds to the limit of $p_c \to 1$;
    and a dashed blue line means that the relation holds only for some special cases (e.g., for Gaussian tasks or when the observation is 1-dimensional).
    \autoref{tab:experiments} summarizes which of these conditions are satisfied in several experimental paradigms used to study measures of surprise.
    Two lines indicate that one of the conditions is sufficient for the two measures to be indistinguishable.
    The text beside each line shows where in the text the existence of the mapping is proven, e.g., R1, C2, and P3 stand for Remark 1, Corollary 2, and Proposition 3, respectively.
    The purple box includes surprise measures that are computed in the parameter ($\Theta\tS{t}$) space, whereas the surprise measures outside of the purple box are computed in the space of observations ($Y\tS{t}$).
See \autoref{sec:surprise_theory} for details.
    \textit{Color should be used in print.}}
    \label{fig:surprise_relation_schematic}
\end{figure*}

Surprise measures are commonly used in experiments to study whether a behavioral or physiological variable $Z$ (e.g., the amplitude of the EEG P300 component \parencite{kolossa2015computational}) is sensitive to or representative of surprise.
Given two measures of surprise $\Ss$ and $\Ss'$, a typical experimental question is which one of them (if any) more accurately explains the variations of the variable $Z$ \parencite{kolossa2015computational, ostwald2012evidence, visalli2021EEG, gijsen2021};
see \autoref{fig:surprise_relation_schematic}A1.
However, if there exists a strictly increasing mapping between $\Ss$ and $\Ss'$ (e.g., as in \autoref{fig:surprise_relation_schematic}A2), then the two surprise measures have the same explanatory power with respect to $Z$ -- because any function of $\Ss$ can be written in terms of $\Ss'$ and vice-versa.
For example, assume that $\Ss = f(\Ss')$ for a strictly increasing function $f$.
If an estimator of the variable $Z$ is found using the measure $\Ss$ as $\hat{Z} = g(\Ss)$, then we can rewrite the same estimator in terms of $\Ss'$ as $\hat{Z} = \Tilde{g}(\Ss') = g(f(\Ss'))$.
Because $g(\Ss)$ and $\Tilde{g}(\Ss')$ have the same explanatory power given any function $g$ and any measure of performance, the two surprise measures $\Ss$ and $\Ss'$ are equally informative about the variable $Z$ in this regard\footnote{
This statement is not necessarily true if one restricts the estimators to a particular class of functions -- e.g., if the estimators are constrained to be linear with respect to surprise measures while $f$ is nonlinear.
Such limitations can be avoided by using non-parametric statistical methods like Spearman or Kendall correlations \parencite{corder2014nonparametric}.
For example, the Spearman correlation (a measure of monotonic relationship between two random variables) between $\Ss'$ and $Z$ is the same as the Spearman correlation between $\Ss = f(\Ss')$ and $Z$, but this is not the case for Pearson correlation (a measure of linear relationship between two random variables) if $f$ is nonlinear.}.
We formalize this idea in Definition \ref{def:indis}.
\begin{mydef}
\label{def:indis}
(Indistinguishability)
For the generative model of Definition \ref{def:gen_model}, we say $\Ss$ and $\Ss'$ are indistinguishable if there exists a strictly increasing function $f: \RR \to \RR$ such that $\Ss = f(\Ss')$ for all choices of belief $\pi\tSb{t}$, cue $x\tS{t}$, and observation $y\tS{t}$.
\end{mydef}
One of our goals in the next three sections is to determine under what conditions different surprise measures are indistinguishable (\autoref{fig:surprise_relation_schematic}B and \autoref{tab:experiments}).

\begin{table*}[t]
\centering
\caption{\label{tab:experiments}
Indistinguishability conditions of \autoref{fig:surprise_relation_schematic} for several experimental paradigms.
Publications specified by $\diamond$ use a generative model similar to ours to describe their experiment from the point of view of participants, even if the actual experimental condition has a slightly different structure compared to their generative model.
Publications specified by $*$ include either (i) features that are not part of our generative model or (ii) additional experiments not covered by our model.
See the original publications for details and \autoref{fig:experiments} for a description of four of the tasks.
A value $p_c>0$ in the last column indicates a volatile environment; however, we note that participants may by default assume that the environment is volatile even in situations where the actual experimental conditions are stationary \parencite{meyniel2016human}.
}
\begin{tabular}{l||l|l|l|l}
    & Task & $\pi\tSb{0}$ & $P(.|x;\pi\tSb{0})$ & $p_c$ \\\hline\hline
\cite{nassar2010approximately,nassar2012rational}$^{\diamond}$   & Volatile Gaussian & = flat & = flat & $>0$ \\\hline
\cite{glaze2015normative}$^{\diamond,*}$  & Volatile 2D Gaussian & = flat & $\neq$ flat  &  $>0$ \\\hline
\cite{o2013dissociable}   & Volatile Gaussian with outliers &  = flat  &  = flat & $>0$ \\
\cite{visalli2021EEG}   &   &  &  &  \\\hline
\cite{squires1976effect}   & Oddball  & = flat & = flat & $=0$ \\
\cite{mars2008trial}$^{\diamond}$  &  &  &  &  \\
\cite{maheu2019brain}$^{\diamond}$, etc.   &  &  &  &  \\\hline
\cite{heilbron2019confidence}$^{\diamond}$  & Volatile oddball  & = flat & = flat & $>0$ \\
\cite{meyniel2020brain}$^{\diamond}$  &  &  &  &  \\\hline
\cite{ostwald2012evidence}$^{\diamond}$   & Roving oddball  & = flat & = flat & $=0$ \\
\cite{lieder2013modelling}  &  &  &  &  \\\hline
\cite{gijsen2021}$^{\diamond}$  & Volatile roving oddball  & = flat & = flat & $>0$ \\\hline
\cite{kolossa2015computational}$^{\diamond}$  & Urn-ball  & $\neq$ flat & $\neq$ flat & $=0$ \\\hline
\cite{behrens2007learning}$^{\diamond}$  & Reversal bandit  & = flat & = flat & $>0$  \\
\cite{horvath2020human}$^{\diamond}$ &  &  &  & \\\hline
\cite{rouhani2021signed}$^*$  & Volatile contextual bandit  & = flat & = flat & $>0$ \\
\cite{findling2019imprecise}$^{\diamond}$  &   &  &  & \\\hline
\cite{glascher2010states}  & Multi-step decision-making & = flat & = flat & $=0$ \\\hline
\cite{liakoni2022brain}$^{\diamond}$   & Multi-step decision-making with outliers & $\neq$ flat & = flat & $=0$\\\hline
\cite{xu2021novelty}$^{\diamond}$   & Volatile multi-step decision-making & $\neq$ flat & = flat & $>0$
\end{tabular}
\end{table*}
\section{Probabilistic mismatch surprise measures}
\label{sec:surprise_theory_prob}
\subsection{Bayes Factor surprise}
\label{sec:surprise_theory_prob_BF}

An abrupt change in the parameters of the environment influences the sequence of observations.
Therefore, a sensible way to define the surprise of an observation is that `surprise' measures the probability of an abrupt change in the eye of the agent, given the present observation.
To detect an abrupt change, it is not enough to measure how unexpected the observation is according to the current belief of the agent.
Rather, the agent should measure how much more expected the new observation is under the prior belief than under the current belief.
The Bayes Factor surprise was introduced by \cite{liakoni2019approximate} to quantify this concept of surprise, motivated by the idea that surprise modulates the speed of learning in the brain \parencite{iigaya2016adaptive, fremaux2016neuromodulated}.

Here, we apply their definition to our generative model.
Similar to \cite{xu2021novelty}, we define the Bayes Factor surprise of observing $y\tS{t+1}$ given the cue $x\tS{t+1}$ as the ratio of the marginal probability of observing $y\tS{t+1}$ given $x\tS{t+1}$ and $C\tS{t+1} = 1$ (i.e., assuming a change) to the marginal probability of observing $y\tS{t+1}$ given $x\tS{t+1}$ and $C\tS{t+1} = 0$ (i.e. assuming no change):
\begin{linenomath*} \begin{equation}\ \begin{aligned}
\label{eq:BF}
\Ss_{\rm BF}(y\tS{t+1}|x\tS{t+1}; \pi\tSb{t}) &\coloneqq \frac{\PP\tSb{t}\big(y\tS{t+1}|x\tS{t+1}, C\tS{t+1} = 1 \big)}{\PP\tSb{t}\big(y\tS{t+1}|x\tS{t+1}, C\tS{t+1} = 0 \big)}\\
&= \frac{P(y\tS{t+1}|x\tS{t+1};\pi\tSb{0})}{P(y\tS{t+1}|x\tS{t+1};\pi\tSb{t})}.
\end{aligned} \end{equation} \end{linenomath*}
The name arises because $\Ss_{\rm BF}(y\tS{t+1}|x\tS{t+1}; \pi\tSb{t})$ is the Bayes Factor \parencite{kass1995bayes, bayarri1997measures} used in statistics to test whether a change has occurred at time $t$.
For a given $P(y\tS{t+1}|x\tS{t+1};\pi\tSb{0})$, the Bayes Factor surprise is a decreasing function of $P(y\tS{t+1}|x\tS{t+1};\pi\tSb{t})$: Hence, more probable events are perceived as less surprising.
However, the key feature of $\Ss_{\rm BF}(y\tS{t+1}|x\tS{t+1}; \pi\tSb{t})$ is that it measures not only how unexpected (unlikely) the observation $y\tS{t+1}$ is according to the current belief $\pi\tSb{t}$ but also how expected it would be if the agent had reset its belief to the prior belief.
More precisely, for a given $P(y\tS{t+1}|x\tS{t+1};\pi\tSb{t})$, the Bayes Factor surprise is an increasing function of $P(y\tS{t+1}|x\tS{t+1};\pi\tSb{0})$.

Such a comparison is necessary to evaluate whether a reset of the belief (or an increase in the update rate of the belief) can be beneficial in order to have a more accurate estimate of the environment's parameters (c.f. \cite{soltani2019adaptive}).
This intuition is formulated in a precise way by \cite{liakoni2019approximate} in their Proposition 1, where they show that, for the generative model of \autoref{fig:gen_model}B, the exact Bayesian inference for the update of $\pi\tSb{t}$ to $\pi\tSb{t+1}$ upon observing $y\tS{t+1}$ leads to a learning rule modulated by the Bayes Factor surprise.
Proposition \ref{Prop:update} below states that this result is also true for our more general generative model (\autoref{fig:gen_model}A).
\begin{myprop}
\label{Prop:update}
(Extension of Proposition 1 of \cite{liakoni2019approximate})
For the generative model of Definition \ref{def:gen_model}, the Bayes Factor surprise can be used to write the updated (according to exact Bayesian inference) belief $\pi\tSb{t+1}$, after observing $y\tS{t+1}$ with the cue $x\tS{t+1}$, as
\begin{linenomath*} \begin{equation}\ \begin{aligned}
\pi\tSb{t+1}(\theta) =
(1-\gamma\tS{t+1}) \pi\tSb{t+1}_{\rm integration}(\theta) +
\gamma\tS{t+1} \pi\tSb{t+1}_{\rm reset}(\theta),
\end{aligned} \end{equation} \end{linenomath*}
where $\gamma\tS{t+1}$ is an adaptation rate modulated by the Bayes Factor surprise
\begin{linenomath*} \begin{equation}\ \begin{aligned}
\gamma\tS{t+1} &\coloneqq \frac{m \Ss_{\rm BF}(y\tS{t+1}|x\tS{t+1}; \pi\tSb{t})}{1 + m\Ss_{\rm BF}(y\tS{t+1}|x\tS{t+1}; \pi\tSb{t})}\\
m &\coloneqq \frac{p_c}{1 - p_c},
\end{aligned} \end{equation} \end{linenomath*}
and
\begin{linenomath*} \begin{equation}\ \begin{aligned}
\pi\tSb{t+1}_{\rm integration}(\theta) &\coloneqq \frac{P_{Y|X}(y\tS{t+1}|x\tS{t+1};\theta) \pi\tSb{t}(\theta)}{P(y\tS{t+1}|x\tS{t+1};\pi\tSb{t})},\\
\pi\tSb{t+1}_{\rm reset}(\theta) &\coloneqq \frac{P_{Y|X}(y\tS{t+1}|x\tS{t+1};\theta) \pi\tSb{0}(\theta)}{P(y\tS{t+1}|x\tS{t+1};\pi\tSb{0})}.
\end{aligned} \end{equation} \end{linenomath*}
\end{myprop}
Therefore, the Bayes Factor surprise $\Ss_{\rm BF}$ controls the trade-off between the integration of the new observation into the old belief (via $\pi\tSb{t+1}_{\rm integration}$) and resetting the old belief to the prior belief (via $\pi\tSb{t+1}_{\rm reset}$).

\subsection{Shannon surprise}

No matter if there has been an abrupt change ($C\tS{t+1} = 1$) or not ($C\tS{t+1} = 0$), an unlikely event may be perceived as surprising.
Therefore, another way to measure the surprise of an observation is to quantify how unlikely the observation is in the eye of the agent.
Shannon surprise, also known as surprisal \parencite{barto2013novelty}, is a way to formalize this concept of surprise.
It comes from the field of information theory \parencite{shannon1948mathematical} and statistical physics \parencite{tribus1961thermostatics} and is widely used in neuroscience \parencite{mars2008trial, kopp2013electrophysiological, kolossa2015computational,  konovalov2018neurocomputational, meyniel2016human, modirshanechi2019trial, maheu2019brain, mousavi2020brain, gijsen2021, visalli2021EEG}.

Formally, for the generative model of Definition \ref{def:gen_model}, one can define the Shannon surprise of observing $y\tS{t+1}$ given the cue $x\tS{t+1}$ as
\begin{linenomath*} \begin{equation}\ \begin{aligned}
\label{eq:Sh1}
\Ss_{\rm Sh1}(y\tS{t+1}|x\tS{t+1}; \pi\tSb{t}) &\coloneqq - \log \PP\tSb{t}\big(y\tS{t+1}|x\tS{t+1} \big) \\
= - \log \Big( &p_c P(y\tS{t+1}|x\tS{t+1};\pi\tSb{0}) + \\
&(1-p_c) P(y\tS{t+1}|x\tS{t+1};\pi\tSb{t})\Big),
\end{aligned} \end{equation} \end{linenomath*}
where the 2nd equality is a result of the marginalization
\begin{linenomath*} \begin{equation}\ \begin{aligned}
\PP\tSb{t}\big(y\tS{t+1}|x\tS{t+1} \big) = \sum_c \PP\tSb{t}\big(y\tS{t+1},C\tS{t+1}=c|x\tS{t+1} \big).
\end{aligned} \end{equation} \end{linenomath*}

The Shannon surprise $\Ss_{\rm Sh1}$ measures how unexpected or unlikely $y\tS{t+1}$ is considering the possibility that there might have been an abrupt change in the environment.
As a result, for a fixed $P(y\tS{t+1}| \allowbreak x\tS{t+1};\pi\tSb{t})$, the Shannon surprise is a decreasing function of $P(y\tS{t+1}|x\tS{t+1};\pi\tSb{0})$ (c.f. \autoref{eq:Sh1}):
It is \textit{less} surprising to observe an event that is more probable under the prior belief because this event is also in total more probable if we consider the possibility of an abrupt change at time $t+1$.
In contrast, the Bayes Factor surprise is an increasing function of $P(y\tS{t+1}|x\tS{t+1};\pi\tSb{0})$ (c.f. \autoref{eq:BF}):
It is \textit{more} surprising to observe an event that is more probable under the prior belief because such events indicate higher chances that an abrupt change has occurred.
This essential difference between the Shannon and the Bayes Factor surprise has been exploited by \cite{liakoni2019approximate} to propose experiments where these two measures of surprise make different predictions.

Experimental evidence \parencite{nassar2010approximately, nassar2012rational} indicates that in volatile environments like the one in \autoref{fig:gen_model}B, human participants do not actively consider the possibility that there \textit{may be} an abrupt change while predicting the \textit{next} observation $y\tS{t+1}$ -- even though they update their belief after observing $y\tS{t+1}$ by considering the possibility that there \textit{might have been} a change before the \textit{current} observation at time $t+1$.
To arrive at a Shannon surprise measure consistent with this observation, we suggest a second definition:
\begin{linenomath*} \begin{equation}\ \begin{aligned}
\label{eq:Sh2}
\Ss_{\rm Sh2}(y\tS{t+1}|x\tS{t+1}; \pi\tSb{t}) &\coloneqq - \log \PP\tSb{t}\big(y\tS{t+1}|x\tS{t+1}, C\tS{t+1}=0 \big) \\
&= - \log P(y\tS{t+1}|x\tS{t+1};\pi\tSb{t}).
\end{aligned} \end{equation} \end{linenomath*}
In other words, $\Ss_{\rm Sh2}(y\tS{t+1}|x\tS{t+1}; \pi\tSb{t})$ neglects the potential presence of change-points, and, therefore, it is independent of both $p_c$ and $P(y\tS{t+1}|x\tS{t+1};\pi\tSb{0})$.
For a non-volatile environment that does not allow for abrupt changes ($p_c = 0$), the two definitions of Shannon surprise are identical: $\Ss_{\rm Sh1} = \Ss_{\rm Sh2}$ (\autoref{fig:surprise_relation_schematic}B).

Proposition \ref{Prop:dSh_BF} shows that the Bayes Factor surprise $\Ss_{\rm BF}$ is related to $\Ss_{\rm Sh1}$ and $\Ss_{\rm Sh2}$:
\begin{myprop}
\label{Prop:dSh_BF}
(Relation between the Shannon surprise and the Bayes Factor surprise)
For the generative model of Definition \ref{def:gen_model},
the Bayes Factor surprise $\Ss_{\rm BF}(y\tS{t+1}|x\tS{t+1}; \pi\tSb{t})$ can be written as
\begin{linenomath*} \begin{equation}\ \begin{aligned}
\Ss_{\rm BF}(y\tS{t+1}|x\tS{t+1}; \pi\tSb{t}) &=
\frac
{(1-p_c)e^{\Delta \Ss_{\rm Sh1}(y\tS{t+1}|x\tS{t+1}; \pi\tSb{t})}}
{1 - p_c e^{\Delta \Ss_{\rm Sh1}(y\tS{t+1}|x\tS{t+1}; \pi\tSb{t})}}\\
&= e^{\Delta \Ss_{\rm Sh2}(y\tS{t+1}|x\tS{t+1}; \pi\tSb{t})},
\end{aligned} \end{equation} \end{linenomath*}
where
\begin{linenomath*} \begin{equation}\ \begin{aligned}
\Delta \Ss_{{\rm Sh}i}(y\tS{t+1}|x\tS{t+1}; \pi\tSb{t}) \coloneqq \, & \Ss_{{\rm Sh}i}(y\tS{t+1}|x\tS{t+1}; \pi\tSb{t}) - \\
&\Ss_{{\rm Sh}i}(y\tS{t+1}|x\tS{t+1}; \pi\tSb{0})
\end{aligned} \end{equation} \end{linenomath*}
for $i \in \{ 1,2\}$.
\end{myprop}
Proposition \ref{Prop:dSh_BF} states that the Bayes Factor $\Ss_{\rm BF}(y\tS{t+1}|\allowbreak x\tS{t+1}; \pi\tSb{t})$ has a behavior similar to the \textit{difference} in Shannon surprise (i.e., $\Delta \Ss_{\rm Sh1}$ or $\Delta \Ss_{\rm Sh2}$) as opposed to Shannon surprise itself (i.e., $\Ss_{\rm Sh1}$ or $\Ss_{\rm Sh2}$).
The difference in Shannon surprise (i.e., $\Delta \Ss_{\rm Sh1}$ or $\Delta \Ss_{\rm Sh2}$) compares the Shannon surprise under the current belief with that under the prior belief.
Two direct consequences of this proposition are summarized in Corollaries \ref{Corr:gamma_Sh} and \ref{Corr:flat_prior}.

Corollary \ref{Corr:gamma_Sh} states that the modulation of learning as presented in Proposition \ref{Prop:update} can also be written in the form of the difference in Shannon surprise (i.e., $\Delta \Ss_{\rm Sh1}$ or $\Delta \Ss_{\rm Sh2}$).

\begin{mycorrol}
\label{Corr:gamma_Sh}
The adaptation rate $\gamma\tS{t+1}$ in Proposition \ref{Prop:update} can be written as
\begin{linenomath*} \begin{equation}\ \begin{aligned}
\gamma\tS{t+1} &= p_c \exp \Big( \Delta \Ss_{\rm Sh1}(y\tS{t+1}|x\tS{t+1}; \pi\tSb{t}) \Big)\\
\gamma\tS{t+1} &= {\rm Sigmoid}\Big(  \Tilde{m} \Delta \Ss_{\rm Sh2}(y\tS{t+1}|x\tS{t+1}; \pi\tSb{t}) \Big),
\end{aligned} \end{equation} \end{linenomath*}
with $\Tilde{m} \coloneqq \log \frac{p_c}{1-p_c} = \log m$ (c.f. Proposition \ref{Prop:update}) and ${\rm Sigmoid}(u) \coloneqq \frac{1}{1 + e^{-u}}$
\end{mycorrol}

Corollary \ref{Corr:flat_prior} indicates that, under a flat prior, the Bayes Factor surprise and the two definitions of the Shannon surprise are indistinguishable from each other (\autoref{fig:surprise_relation_schematic}B):
\begin{mycorrol}
\label{Corr:flat_prior}
(Flat prior prediction)
For the generative model of Definition \ref{def:gen_model}, if the probability of observing $y\tS{t+1}$ with the cue $x\tS{t+1}$ is flat under the prior belief $\pi\tSb{0}$ (i.e., if $P(y\tS{t+1}|x\tS{t+1};\pi\tSb{0})$ is uniform), then there are strictly increasing mappings between $\Ss_{\rm BF}(y\tS{t+1}| \allowbreak x\tS{t+1};\pi\tSb{t})$, $\Ss_{\rm Sh1}(y\tS{t+1}|\allowbreak x\tS{t+1}; \pi\tSb{t})$, and $\Ss_{\rm Sh2}(y\tS{t+1}| x\tS{t+1}; \allowbreak \pi\tSb{t})$.
\end{mycorrol}

A consequence of Corollary \ref{Corr:flat_prior} is that experiments with flat marginal priors of the agent cannot be used to distinguish $\Ss_{\rm BF}$ from $\Ss_{\rm Sh1}$ or $\Ss_{\rm Sh2}$ (\autoref{fig:surprise_relation_schematic}).

\subsection{State prediction error}

The State Prediction Error (SPE) was introduced by \cite{glascher2010states} in the context of model-based reinforcement learning in Markov Decision Processes (MDPs -- c.f. \autoref{fig:gen_model}E) \parencite{sutton2018reinforcement}.
Similar to the Shannon surprise, the SPE considers less probable events as the more surprising ones.

Whenever observations $y\tS{1:t}$ come from a discrete distribution so that we have $P_{Y|X}(y\tS{t+1}|x\tS{t+1};\allowbreak \theta) \in [0,1]$ for all $\theta$, $x\tS{t+1}$, and $y\tS{t+1}$, we can generalize the definition of \cite{glascher2010states} to the setting of our generative model.
Analogously to our two definitions of Shannon surprise (c.f. \autoref{eq:Sh1} and \autoref{eq:Sh2}), we give also two definitions for SPE:
\begin{linenomath*} \begin{equation}\ \begin{aligned}
\label{eq:SPE1}
\Ss_{\rm SPE1}(y\tS{t+1}|x\tS{t+1}; \pi\tSb{t}) &\coloneqq 1 - \PP\tSb{t}\big(y\tS{t+1}|x\tS{t+1} \big) \\
= 1 - \Big( &p_c P(y\tS{t+1}|x\tS{t+1};\pi\tSb{0}) + \\
            &(1-p_c) P(y\tS{t+1}|x\tS{t+1};\pi\tSb{t})\Big),
\end{aligned} \end{equation} \end{linenomath*}
and
\begin{linenomath*} \begin{equation}\ \begin{aligned}
\label{eq:SPE2}
\Ss_{\rm SPE2}(y\tS{t+1}|x\tS{t+1};& \pi\tSb{t})\\
\coloneqq &1 - \PP\tSb{t}\big(y\tS{t+1}|x\tS{t+1}, C\tS{t+1}=0 \big) \\
= &1 - P(y\tS{t+1}|x\tS{t+1};\pi\tSb{t}).
\end{aligned} \end{equation} \end{linenomath*}
In non-volatile environments ($p_c = 0$), the two definitions of SPE are identical (\autoref{fig:surprise_relation_schematic}B).
In particular, in an MDP without abrupt changes ($p_c = 0$; \autoref{fig:gen_model}E), both definitions are equal to $1 - \PP^{(t)}(s_t,a_t \to s_{t+1})$, where $\PP^{(t)}(s_t,a_t \to s_{t+1})$ is an agent's estimate (at time $t$) of the probability of the transition to state $s_{t+1}$ after taking action $a_t$ in state $s_t$; c.f. \cite{glascher2010states}.

Proposition \ref{Prop:SPE_Sh} states that both definitions ($\Ss_{\rm SPE1}$ and $\Ss_{\rm SPE2}$) can always be written as strictly increasing functions of Shannon surprise (\autoref{fig:surprise_relation_schematic}B):

\begin{myprop}
\label{Prop:SPE_Sh}
(Relation between the Shannon surprise and the SPE)
For the generative model of Definition \ref{def:gen_model}, for $i \in \{1,2\}$, the state prediction error $\Ss_{{\rm SPE}i}(y\tS{t+1}|x\tS{t+1}; \pi\tSb{t})$, can be written as
\begin{linenomath*} \begin{equation}\ \begin{aligned}
\Ss_{{\rm SPE}i}(y\tS{t+1}&|x\tS{t+1}; \pi\tSb{t}) = \\
&1 - \exp \Big( - \Ss_{{\rm Sh}i}(y\tS{t+1}|x\tS{t+1}; \pi\tSb{t})  \Big).
\end{aligned} \end{equation} \end{linenomath*}
\end{myprop}

Therefore, the SPE and the Shannon surprise are indistinguishable (\autoref{fig:surprise_relation_schematic}).

\section{Observation-mismatch surprise measures}
\label{sec:surprise_theory_obs}
\subsection{Absolute and squared errors}

Assume an agent predicts $\hat{y}\tS{t+1}$ for the next observation $y\tS{t+1}$. Then, a measure of surprise can be defined as the prediction error or the mismatch between the prediction $\hat{y}\tS{t+1}$ and the actual observation $y\tS{t+1}$ \parencite{nassar2010approximately, nassar2012rational, prat2021human} (\autoref{fig:surprise_belief}).
For the sake of completeness, we discuss four possible definitions for observation-mismatch surprise measures.

Before turning to an `observation-mismatch', we first need to define an agent's prediction for the next observation.
Analogously to our two definitions for the Shannon surprise (c.f. \autoref{eq:Sh1} and \autoref{eq:Sh2}), we define two different predictions for the next observation $y\tS{t+1}$ given the cue $x\tS{t+1}$\footnote{The evaluation of the full distribution $P(.|x\tS{t+1};\pi\tSb{t})$ may not always be necessary for the computation of $E_1$ and $E_2$ \parencite{nassar2010approximately, liakoni2019approximate, aguilera2021particular}.}:
\begin{linenomath*} \begin{equation}\ \begin{aligned}
\label{eq:E1}
E_1 [Y\tS{t+1}] \coloneqq
&p_c \EE_{P(.|x\tS{t+1};\pi\tSb{0})} [ Y\tS{t+1} ] +\\
&(1-p_c) \EE_{P(.|x\tS{t+1};\pi\tSb{t})} [ Y\tS{t+1} ]
\end{aligned} \end{equation} \end{linenomath*}
and
\begin{linenomath*} \begin{equation}\ \begin{aligned}
\label{eq:E2}
E_2 [Y\tS{t+1}] \coloneqq \EE_{P(.|x\tS{t+1};\pi\tSb{t})} [ Y\tS{t+1} ].
\end{aligned} \end{equation} \end{linenomath*}
Although $E_1 [Y\tS{t+1}]$ is a more reasonable prediction for $y\tS{t+1}$ given the fact that there is always a possibility of an abrupt change according to our generative model of the environment (Definition \ref{def:gen_model}), \cite{nassar2010approximately} have shown that, in a Gaussian task, $E_2 [Y\tS{t+1}]$ explains human participants' predictions better than $E_1 [Y\tS{t+1}]$.

We note that the observation $y\tS{t+1}$ is, in general, multi-dimensional.
As two natural ways of measuring mismatch, we define the squared and the absolute error surprise, for $i \in \{1,2\}$, as
\begin{linenomath*} \begin{equation}\ \begin{aligned}
\label{eq:Ab_Sq}
\Ss_{{\rm Ab},i}(y\tS{t+1}|x\tS{t+1}; \pi\tSb{t}) &\coloneqq ||y\tS{t+1} -  E_i [Y\tS{t+1}]||_1\\
\Ss_{{\rm Sq},i}(y\tS{t+1}|x\tS{t+1}; \pi\tSb{t}) &\coloneqq \Big(||y\tS{t+1} -  E_i [Y\tS{t+1}]||_2\Big)^2,
\end{aligned} \end{equation} \end{linenomath*}
where $||.||_1$ and $||.||_2$ stand for the $\ell_1$- and $\ell_2$-norms (c.f. \autoref{tab:notation}), respectively, and $E_1$ and $E_2$ are defined in \autoref{eq:E1} and \autoref{eq:E2}, respectively.
Similar definitions have been used in neuroscience \parencite{nassar2010approximately, prat2021human} and machine learning \parencite{pathak2017curiosity, burda2018large}.
In Propositions \ref{Prop:Ab_Sq_Cat}-\ref{Prop:Ab_Sq}, we show for three special cases that the absolute and the squared error surprise can be written as strictly increasing functions of either each other or the SPE and the Shannon surprise (\autoref{fig:surprise_relation_schematic}B).
\begin{myprop}
\label{Prop:Ab_Sq_Cat}
(Relation between the absolute and squared errors and the SPE for categorical distributions)
For the generative model of Definition \ref{def:gen_model},
if $Y\tS{t+1}$ is represented as one-hot coded vectors, i.e., vectors with one element equal to 1 and the others equal to 0, then we have, for $i \in \{1,2\}$,
\begin{linenomath*} \begin{equation}\ \begin{aligned}
\Ss_{{\rm Ab}i}(y\tS{t+1}|x\tS{t+1}; \pi\tSb{t}) = 2 \Ss_{{\rm SPE}i}(y\tS{t+1}|x\tS{t+1}; \pi\tSb{t}),
\end{aligned} \end{equation} \end{linenomath*}
and
\begin{linenomath*} \begin{equation}\ \begin{aligned}
\Ss_{{\rm Sq}i}(y\tS{t+1}|x\tS{t+1}; \pi\tSb{t}) = &2 \Ss_{{\rm SPE}i}(y\tS{t+1}|x\tS{t+1}; \pi\tSb{t}) + \\
& {\rm Conf.}\Big[P(.|x\tS{t+1};\pi\tSb{t})  \Big],
\end{aligned} \end{equation} \end{linenomath*}
where ${\rm Conf.}\Big[P(.|x\tS{t+1};\pi\tSb{t}) \Big]$ can be seen as a measure of confidence in the prediction (see \ref{sec:app_proofs}).
\end{myprop}
\begin{myprop}
\label{Prop:Sq_Gau}
(Relation between the squared error surprise and the Shannon surprise for Gaussian distributions -- from \cite{pathak2017curiosity})
For the generative model of Definition \ref{def:gen_model},
if the marginal distribution of $Y\tS{t+1} \in \RR^N$ given the cue $x\tS{t+1}$ and the belief $\pi\tSb{t}$ is a Gaussian distribution with a covariance matrix equal to $\sigma I_{N \times N}$, where $I_{N \times N}$ is the $N\times N$ identity matrix, then $\Ss_{\rm Sq2}(y\tS{t+1}|x\tS{t+1}; \allowbreak \pi\tSb{t})$ is a strictly increasing function of $\Ss_{\rm Sh2}(y\tS{t+1}|x\tS{t+1}; \pi\tSb{t})$.
\end{myprop}
\begin{myprop}
\label{Prop:Ab_Sq}
(Observation-mismatch surprise measures for 1-D observations)
For the generative model of Definition \ref{def:gen_model}, if $Y\tS{t} \in \RR$, then we have $\Ss_{{\rm Sq}i} = \Ss_{{\rm Ab}i}^2$ for $i \in \{1,2\}$ implying that the two observation-mismatch surprise measures are indistinguishable.
\end{myprop}
We note that, according to Proposition \ref{Prop:SPE_Sh}, the SPE is a strictly increasing function of the Shannon surprise.
Hence, for categorical distributions with one-hot coding, the SPE, the Shannon surprise, and the absolute error surprise are indistinguishable, and for Gaussian distributions with scaled identity covariance, the SPE, the Shannon surprise, and the squared error surprise are indistinguishable (\autoref{fig:surprise_relation_schematic}).

\subsection{Unsigned reward prediction error}

A particular form of observation-mismatch surprise in the context of reward-driven decision making is the Unsigned Reward Prediction Error (uRPE, i.e., the absolute value of Reward Prediction Error) \parencite{pearce1980model, hayden2011surprise, talmi2013feedback, roesch2012surprise, rouhani2021signed}.
In this section, we first discuss the definition of the uRPE as it often appears in experimental studies and then analyze a generalized definition of the uRPE in general sequential decision-making tasks.

Many of the experimental paradigms (e.g., \cite{hayden2011surprise, talmi2013feedback, roesch2012surprise}) for the study of uRPE can be modeled by a non-volatile (i.e., $p_c = 0$) contextual bandit task where, given a context $s\tS{t}$ (e.g., conditioned stimulus), the agent takes an action $a\tS{t}$ and receives a real-valued reward $r\tS{t+1}$.
The uRPE corresponding to the tuple $(s\tS{t},a\tS{t},r\tS{t+1})$ is \parencite{sutton2018reinforcement}
\begin{linenomath*} \begin{equation}\ \begin{aligned}
\label{eq:uRPE_simp}
{\rm uRPE}(s\tS{t},a\tS{t} \to r\tS{t+1}) \coloneqq | r\tS{t+1} - Q\tSb{t}(s\tS{t},a\tS{t}) |,
\end{aligned} \end{equation} \end{linenomath*}
where $Q\tSb{t}(s\tS{t},a\tS{t})$ is the latest estimate of the expectation of $R\tS{t+1}$ given $s\tS{t}$ and $a\tS{t}$.
The generative model of Definition \ref{def:gen_model} is reduced to a model of contextual bandit tasks if we put $X\tS{t+1} \coloneqq (S\tS{t},A\tS{t})$ and $Y\tS{t+1} \coloneqq R\tS{t+1}$.
Then, the unsigned reward prediction error ${\rm uRPE}(s\tS{t},a\tS{t} \to r\tS{t+1})$ is syntactically equal to $\Ss_{{\rm Ab}}$ (c.f. \autoref{eq:Ab_Sq}; note that $E_1 = E_2$ since $p_c = 0$) and indistinguishable from $\Ss_{{\rm Sq}}$ (Proposition \ref{Prop:Ab_Sq}):
\begin{myremark}
(Relation between the common definition of uRPE and the other two observation-mismatch surprise measures)
The uRPE signal that was previously investigated in many experimental studies (\autoref{eq:uRPE_simp}) \parencite{pearce1980model, hayden2011surprise, talmi2013feedback, roesch2012surprise} is a special case of the absolute and the squared error surprise (\autoref{eq:Ab_Sq}).
\end{myremark}

However, one can go beyond contextual bandit tasks and define uRPE for a general Markov Decision Process (MDP) \parencite{sutton2018reinforcement}.
To reduce our generative model of Definition \ref{def:gen_model} to a (potentially volatile, i.e., $p_c \geq 0$) MDP, we put the cue variable $X\tS{t+1}$ equal to the state-action pair $(S\tS{t},A\tS{t})$ and the observation $Y\tS{t+1}$ equal to the pair of the next state $S\tS{t+1}$ and the next \textit{extended reward} $\Tilde{R}\tS{t+1}$ that we define as
\begin{linenomath*} \begin{equation}\ \begin{aligned}
\label{eq:extended_R}
\Tilde{R}\tS{t+1} \coloneqq R\tS{t+1} + \lambda V(S\tS{t+1}),
\end{aligned} \end{equation} \end{linenomath*}
where $\lambda \in [0,1)$ is the discount factor in infinite-horizon reinforcement learning \parencite{sutton2018reinforcement}, and $V(S\tS{t+1})$ is the \textit{perceived value} of state $S\tS{t+1}$.
Here, we do not discuss the exact definition of $V$ and how it is computed;
we only assume that each state $s$ has a value $V(s)$ that is informative about the expected amount of total reward that one can collect starting from state $s$ -- see \cite{sutton2018reinforcement} for details.
Analogously to our two definitions for the absolute and the squared error surprise (c.f. \autoref{eq:Ab_Sq}), we give two definitions of uRPE:
\begin{linenomath*} \begin{equation}\ \begin{aligned}
\label{eq:uRPE_comp}
\Ss_{{\rm uRPE}i}(y\tS{t+1}|x\tS{t+1}; &\pi\tSb{t}) \coloneqq\\
&| r\tS{t+1} + \lambda V(s\tS{t+1})  - Q_i\tSb{t}(s\tS{t},a\tS{t}) |,
\end{aligned} \end{equation} \end{linenomath*}
where $i \in \{1,2\}$ and $Q_i\tSb{t}(s\tS{t},a\tS{t}) \coloneqq E_i [\Tilde{R}\tS{t+1}]$ (c.f. \autoref{eq:E1}, \autoref{eq:E2}, and \autoref{eq:extended_R}).
\autoref{eq:uRPE_comp} implies that the uRPE surprise is like the absolute error surprise if an agent focuses exclusively on the extended reward $\Tilde{r}\tS{t+1}$ and ignores the state $s\tS{t+1}$.
We make this intuition formal in Proposition \ref{Prop:uRPE}.
\begin{myprop}
\label{Prop:uRPE}
(Relation between the uRPE, the absolute error, and squared error surprise measures)
For the generative model of Definition \ref{def:gen_model}, for $i \in \{1,2\}$, the unsigned reward prediction error $\Ss_{{\rm uRPE}i}(y\tS{t+1}| \allowbreak x\tS{t+1}; \pi\tSb{t})$ can be written as
\begin{linenomath*} \begin{equation}\ \begin{aligned}
\Ss_{{\rm uRPE}i}(y\tS{t+1}|x\tS{t+1}; \pi\tSb{t}) = \, &\Ss_{{\rm Ab}i}(y\tS{t+1}|x\tS{t+1}; \pi\tSb{t}) - \\
&\Ss_{{\rm Ab}i}(s\tS{t+1}|x\tS{t+1}; \pi\tSb{t})
\end{aligned} \end{equation} \end{linenomath*}
and
\begin{linenomath*} \begin{equation}\ \begin{aligned}
\Big(\Ss_{{\rm uRPE}i}(y\tS{t+1}|x\tS{t+1}; \pi\tSb{t})\Big)^2 = \, &\Ss_{{\rm Sq}i}(y\tS{t+1}|x\tS{t+1}; \pi\tSb{t}) - \\
&\Ss_{{\rm Sq}i}(s\tS{t+1}|x\tS{t+1}; \pi\tSb{t}).
\end{aligned} \end{equation} \end{linenomath*}
where $\Ss_{{\rm Ab}i}(s\tS{t+1}|x\tS{t+1}; \pi\tSb{t}) \coloneqq ||s\tS{t+1} -  E_i [S\tS{t+1}]||_1$ and $\allowbreak\Ss_{{\rm Sq}i}(s\tS{t+1}|x\tS{t+1}; \pi\tSb{t}) \coloneqq ||s\tS{t+1} -  E_i [S\tS{t+1}]||^2_2$ (\autoref{eq:Ab_Sq}).
\end{myprop}
Therefore, if observation $y\tS{t+1}$ does not include state $s\tS{t+1}$ (e.g., in contextual bandit tasks, similar to \cite{hayden2011surprise, talmi2013feedback, roesch2012surprise}) or if all possible values of state $s\tS{t+1}$ are equally surprising (i.e., have constant $\Ss_{{\rm Sq}i}$ or $\Ss_{{\rm Ab}i}$, similar to the experiment of \cite{rouhani2021signed}), then $\Ss_{{\rm uRPE}i}$ is indistinguishable from $\Ss_{{\rm Ab}i}$ and $\Ss_{{\rm Sq}i}$ (\autoref{fig:surprise_relation_schematic}).

\section{Belief-mismatch surprise measures}
\label{sec:surprise_theory_belief}

\subsection{Bayesian surprise}

Another way to think about surprise is to define surprising events as those that change an agent's belief about the world.
Bayesian surprise \parencite{baldi2002computational, schmidhuber2010formal, baldi2010bits} is a way to formalize this concept of surprise.
Whereas the Bayes Factor surprise measures how likely it is that the environment has changed given the new observation, the Bayesian surprise measures how much the agent's belief changes given the new observation.

Bayesian surprise \parencite{baldi2002computational} has been originally introduced in non-volatile environments, i.e., where there is no change ($p_c=0$) and as a result $\Theta\tS{1} = \Theta\tS{2} = ... = \Theta\tS{t} = \Theta$.
In this case, the Bayesian surprise of observing $y\tS{t+1}$ with cue $x\tS{t+1}$ is defined as $\DKL [\PP\tSb{t}_{\Theta} || \PP\tSb{t+1}_{\Theta}]$ \parencite{baldi2002computational, baldi2010bits, schmidhuber2010formal}, where $\DKL$ stands for the Kullback-Leibler (KL) divergence \parencite{cover1999elements}, and $\PP\tSb{t}_{\Theta}$ is an alternative notation for the distribution of $\Theta$ conditioned on $x\tS{1:t}$ and $y\tS{1:t}$ (c.f. \autoref{tab:notation}).
Hence, in non-volatile environments, Bayesian surprise measures the pseudo-distance $\DKL$ between two distributions, i.e., the belief $\pi\tSb{t} = \PP\tSb{t}_{\Theta}$ before and the belief $\pi\tSb{t+1} = \PP\tSb{t+1}_{\Theta}$ after observing $y\tS{t+1}$.
To generalize this definition to volatile environments, we have to choose two equivalent distributions that we want to compare.
The natural choice for $\PP\tSb{t+1}_{\Theta}$ is $\PP\tSb{t+1}_{\Theta\tS{t+1}} = \pi\tSb{t+1}$;
however, it is unclear whether $\PP\tSb{t}_{\Theta}$ should be taken as the momentary belief $\PP\tSb{t}_{\Theta\tS{t}} = \pi\tSb{t}$ or its one-step forward-propagation $\PP\tSb{t}_{\Theta\tS{t+1}}$ \textit{before} the next observation $y\tS{t+1}$ is integrated.
If $p_c \neq 0$, the two choices are different:
\begin{linenomath*} \begin{equation}\ \begin{aligned}
\label{eq:pi_t_weighted}
\pi\tSb{t} = \PP\tSb{t}_{\Theta\tS{t}} \neq \PP\tSb{t}_{\Theta\tS{t+1}} = p_c \pi\tSb{0} + (1-p_c)\pi\tSb{t}.
\end{aligned} \end{equation} \end{linenomath*}
Therefore, for the case of volatile environments, we give two definitions for the Bayesian surprise:
\begin{linenomath*} \begin{equation}\ \begin{aligned}
\label{eq:Ba1}
\Ss_{\rm Ba1}(y\tS{t+1}|&x\tS{t+1}; \pi\tSb{t}) \coloneqq\\
&\DKL \Big[ p_c \pi\tSb{0} + (1-p_c)\pi\tSb{t} || \pi\tSb{t+1}  \Big],
\end{aligned} \end{equation} \end{linenomath*}
and
\begin{linenomath*} \begin{equation}\ \begin{aligned}
\label{eq:Ba2}
\Ss_{\rm Ba2}(y\tS{t+1}|x\tS{t+1}; \pi\tSb{t}) &\coloneqq \DKL \Big[ \pi\tSb{t} || \pi\tSb{t+1}  \Big].
\end{aligned} \end{equation} \end{linenomath*}
The first definition is more consistent with the original definition of the Bayesian surprise \parencite{baldi2002computational, baldi2010bits, schmidhuber2010formal} applied to our generative model because the belief before the observation should include the knowledge that the environment is volatile.
However, the second definition looks more intuitive from the neuroscience perspective \parencite{gijsen2021, mousavi2020brain}.
Note that, in \autoref{eq:Ba1} and \autoref{eq:Ba2}, the observation $y\tS{t+1}$ does not appear explicitly on the right hand side;
the observation has, however, influenced the update of the belief to its new distribution $\pi\tSb{t+1}$.
For the case of $p_c = 0$, the two definitions are identical (\autoref{fig:surprise_relation_schematic}B).

In Proposition \ref{Prop:Ba_Sh} and Remark \ref{Rem:Ba_Sh}, we show that the Bayesian surprise is correlated with the difference between the Shannon surprise and its expectation (over all possible values of $\Theta\tS{t+1}$).
\begin{myprop}
\label{Prop:Ba_Sh}
(Relation between the Bayesian surprise and the Shannon surprise)
In the generative model of Definition \ref{def:gen_model},
the Bayesian surprise can be written as
\begin{linenomath*} \begin{equation}\
\begin{aligned}
\Ss_{\rm Ba1}(y\tS{t+1}|&x\tS{t+1}; \pi\tSb{t}) = \\
&p_c \EE_{\pi\tSb{0}} \Big[ \Ss_{\rm Sh2}(y\tS{t+1}|x\tS{t+1}; \delta_{\{\Theta\}}) \Big] + \\
&(1-p_c) \EE_{\pi\tSb{t}} \Big[ \Ss_{\rm Sh2}(y\tS{t+1}|x\tS{t+1}; \delta_{\{\Theta\}}) \Big] - \\
& \Ss_{\rm Sh1}(y\tS{t+1}|x\tS{t+1}; \pi\tSb{t}),
\end{aligned}
\end{equation} \end{linenomath*}
and
\begin{linenomath*} \begin{equation}\
\begin{aligned}
\Ss_{\rm Ba2}(y\tS{t+1}|&x\tS{t+1}; \pi\tSb{t}) = \\
&\EE_{\pi\tSb{t}} \Big[ \Ss_{\rm Sh2}(y\tS{t+1}|x\tS{t+1}; \delta_{\{\Theta\}}) \Big] - \\
&\Ss_{\rm Sh1}(y\tS{t+1}|x\tS{t+1}; \pi\tSb{t}) +\\
&\DKL \Big[ \pi\tSb{t} || p_c \pi\tSb{0} + (1-p_c)\pi\tSb{t} \Big],
\end{aligned}
\end{equation} \end{linenomath*}
where $\delta_{\{\theta\}}$ is the Dirac measure at $\theta$ (c.f. \autoref{tab:notation}).
\end{myprop}

\begin{myremark}
\label{Rem:Ba_Sh}
As a direct consequence of Proposition \ref{Prop:Ba_Sh}, when the change point probability is zero, i.e. $p_c=0$, the Bayesian surprise is equal to the expected Shannon surprise minus the Shannon surprise, i.e.,
\begin{linenomath*} \begin{equation}\
\begin{aligned}
\Ss_{\rm Ba}(y\tS{t+1}|x\tS{t+1}; \pi\tSb{t}) = & \, \EE_{\pi\tSb{t}} \Big[ \Ss_{\rm Sh}(y\tS{t+1}|x\tS{t+1}; \delta_{\{\Theta\}}) \Big] -\\
&\Ss_{\rm Sh}(y\tS{t+1}|x\tS{t+1}; \pi\tSb{t}),
\end{aligned}
\end{equation} \end{linenomath*}
where $\Ss_{\rm Ba} = \Ss_{\rm Ba1} = \Ss_{\rm Ba2}$ and $\Ss_{\rm Sh} = \Ss_{\rm Sh1} = \Ss_{\rm Sh2}$.
\end{myremark}

There are two consequences of this observation.
First, Bayesian surprise is distinguishable from Shannon surprise since it cannot be found only as a function of Shannon surprise.
Second, we need access to the full belief distribution $\pi\tSb{t}$ for computing the expectation (\autoref{fig:surprise_belief}).

In general, surprise measures similar to the Bayesian surprise can be defined also by measuring the change in the belief via distance or pseudo-distance measures different from the KL-divergence \parencite{baldi2002computational}.

\subsection{Postdictive surprise}

We saw that the Bayesian surprise measures how much the new belief $\pi\tSb{t+1}$ has changed after observing $y\tS{t+1}$.
\cite{kolossa2015computational} introduced `postdictive surprise' with a similar idea in mind but focused on changes in the marginal distribution $P(y|x\tS{t+1}; \pi\tSb{t+1})$ (c.f. \autoref{eq:marginal_prob}).
More precisely, whereas the Bayesian surprise measures the amount of update in the space of distributions over the parameters (i.e., how differently the agent thinks about the parameters), the postdictive surprise measures the amount of update in the space of distributions over the observations (i.e., how differently the agent predicts the next observations).

Analogous to our two definitions for the Bayesian surprise (\autoref{eq:Ba1} and \autoref{eq:Ba2}), there are two definitions for the postdictive surprise in volatile environments:
\begin{linenomath*} \begin{equation}\ \begin{aligned}
\label{eq:Po1}
\Ss_{\rm Po1}&(y\tS{t+1}|x\tS{t+1}; \pi\tSb{t}) \coloneqq\\
D_{\rm KL} \Big[
&p_c P\big(.|x\tS{t+1}; \pi\tSb{0} \big) + (1-p_c) P\big(.|x\tS{t+1}; \pi\tSb{t} \big) || \\
&P\big(.|x\tS{t+1}; \pi\tSb{t+1} \big)  \Big],
\end{aligned} \end{equation} \end{linenomath*}
and
\begin{linenomath*} \begin{equation}\ \begin{aligned}
\label{eq:Po2}
\Ss_{\rm Po2}(y\tS{t+1}|x\tS{t+1}; \pi\tSb{t}) \coloneqq
D_{\rm KL} \Big[ &P\big(.|x\tS{t+1}; \pi\tSb{t} \big) ||\\
&P\big(.|x\tS{t+1}; \pi\tSb{t+1} \big)  \Big],
\end{aligned} \end{equation} \end{linenomath*}
where the dot refers to a dummy variable $y$ that is integrated out when evaluating $\DKL$ (c.f. \autoref{tab:notation}).
Note that for $p_c=0$, the two definitions are identical (\autoref{fig:surprise_relation_schematic}B).

Although the amount of update is computed over the space of observations, $\Ss_{\rm Po1}$ and $\Ss_{\rm Po2}$ cannot be categorized as probabilistic mismatch surprise measures, since the update depends explicitly on the belief $\pi\tSb{t}$.
The statement is further explained in our Lemma \ref{Lemma:Po_Sh} in \ref{sec:app_proofs}.

\subsection{Confidence Corrected surprise}

Since surprise arises when an expectation is violated, the violation of an agent's expectation should be more surprising when the agent is more confident about its expectation.
Based on the observation that neither Shannon nor Bayesian surprise explicitly captures the concept of confidence, \cite{faraji2018balancing} proposed the `Confidence Corrected Surprise' as a new measure of surprise that explicitly takes confidence into account.

To define the Confidence Corrected surprise, we first define $\pi_\flatp$ as the flat (uniform) distribution over the space of parameters, i.e., over the set to which $\Theta\tS{t}$ belongs.
Then, following \cite{faraji2018balancing}, we define the normalized likelihood after observing $y\tS{t+1}$ (i.e., the posterior given the flat prior) as
\begin{linenomath*} \begin{equation}\ \begin{aligned}
\pi_\flatp(\theta|y\tS{t+1},x\tS{t+1}) &\coloneqq \frac{P_{Y|X}(y\tS{t+1}|x\tS{t+1};\theta) \pi_\flatp(\theta)}{P(y\tS{t+1}|x\tS{t+1};\pi_\flatp)}\\
&= \frac{P_{Y|X}(y\tS{t+1}|x\tS{t+1};\theta)}{\int P_{Y|X}(y\tS{t+1}|x\tS{t+1};\theta) d \theta}.
\end{aligned} \end{equation} \end{linenomath*}
If the prior $\pi\tSb{0}$ is equal to $\pi_\flatp$ (i.e., if the prior is uniform), then $\pi_\flatp(\theta|y\tS{t+1},x\tS{t+1})$ is the same as $\pi\tSb{t+1}_{\rm reset}(\theta)$ defined in Proposition \ref{Prop:update}.
Note that the prior $\pi_\flatp$ does not necessarily need to be a proper distribution (i.e., does not necessarily need to be normalized) as long as $\int P_{Y|X}(y\tS{t+1}|\allowbreak x\tS{t+1};\theta) d \theta$ is finite and the posterior $\pi_\flatp(.|y\tS{t+1},x\tS{t+1})$ is a proper distribution \parencite{efron2016computer}.
Using this terminology, the original definition for the Confidence Corrected surprise is \parencite{faraji2018balancing}
\begin{linenomath*} \begin{equation}\ \begin{aligned}
\label{eq:CC1}
\Ss_{\rm CC1}(y\tS{t+1}|x\tS{t+1}; \pi\tSb{t}) &\coloneqq \DKL \Big[ \pi\tSb{t} || \pi_\flatp(.|y\tS{t+1},x\tS{t+1}) \Big].
\end{aligned} \end{equation} \end{linenomath*}
To interpret $\Ss_{\rm CC1}$, \cite{faraji2018balancing} defined the commitment (or confidence) $C [ \pi ]$ corresponding to an arbitrary belief $\pi$ as its negative entropy \parencite{cover1999elements}, i.e.,
\begin{linenomath*} \begin{equation}\ \begin{aligned}
\label{eq:faraji_conf}
C [ \pi ] \coloneqq \EE_{\pi} \Big[ \log \pi(\Theta) \Big].
\end{aligned} \end{equation} \end{linenomath*}
Then, in a non-volatile environment (i.e., $p_c = 0$), they show that $\Ss_{\rm CC1}$ can be written as \parencite{faraji2018balancing}
\begin{linenomath*} \begin{equation}\ \begin{aligned}
\label{eq:CC1_Sh_Ba}
\Ss_{\rm CC1}(y\tS{t+1}|x\tS{t+1}; \pi\tSb{t}) =&
\Ss_{\rm Sh}(y\tS{t+1}|x\tS{t+1}; \pi\tSb{t}) +\\
&\Ss_{\rm Ba}(y\tS{t+1}|x\tS{t+1}; \pi\tSb{t}) + \\
&C \big[ \pi\tSb{t} \big] - A (y\tS{t+1},x\tS{t+1}),
\end{aligned} \end{equation} \end{linenomath*}
where $A (y\tS{t+1},x\tS{t+1}) \coloneqq \Ss_{\rm Sh}(y\tS{t+1}|x\tS{t+1}; \pi_\flatp) + C [ \pi_\flatp ]$ is independent of the current belief $\pi\tSb{t}$.
Note that because $p_c=0$, we have $\Ss_{\rm Sh1} = \Ss_{\rm Sh2}$ and $\Ss_{\rm Ba1} = \Ss_{\rm Ba2}$.
Therefore, in a non-volatile environment (i.e., $p_c = 0$), $\Ss_{\rm CC1}$ is correlated with the sum of the Shannon and the Bayesian surprise regularized by the confidence of the agent's belief.
However, such an interpretation is no longer possible in volatile environments ($p_c > 0$), and \autoref{eq:CC1_Sh_Ba} must be replaced by Proposition \ref{Prop:CC} below.

In order to account for the information of the true prior $\pi\tSb{0}$ and to avoid cases where $\pi_\flatp(.|y\tS{t+1},x\tS{t+1})$ is not a proper distribution, we also give a 2nd definition for the Confidence Corrected surprise as
\begin{linenomath*} \begin{equation}\ \begin{aligned}
\label{eq:CC2}
\Ss_{\rm CC2}(y\tS{t+1}|x\tS{t+1}; \pi\tSb{t}) &\coloneqq \DKL \Big[ \pi\tSb{t} || \pi\tSb{t+1}_{\rm reset} \Big],
\end{aligned} \end{equation} \end{linenomath*}
where $\pi\tSb{t+1}_{\rm reset}(\theta)$ is defined in Proposition \ref{Prop:update}.
Whenever $\pi\tSb{0} = \pi_\flatp$, the two definitions are identical (\autoref{fig:gen_model}B).
Proposition \ref{Prop:CC} shows how the Confidence Corrected surprise relates to the Shannon surprise, the Bayesian surprise, and the confidence in the general case.

\begin{myprop}
\label{Prop:CC}
(Relation between the Confidence Corrected surprise, Shannon surprise, and Bayesian surprise)
For the generative model of Definition \ref{def:gen_model}, the original definition of the Confidence Corrected surprise can be written as
\begin{linenomath*} \begin{equation}\ \begin{aligned}
\label{eq:prop_CC1}
\Ss_{\rm CC1}&(y\tS{t+1}|x\tS{t+1}; \pi\tSb{t}) = \\
& \Ss_{\rm Sh1}(y\tS{t+1}|x\tS{t+1}; \pi\tSb{t}) - \Ss_{\rm Sh2}(y\tS{t+1}|x\tS{t+1}; \pi_\flatp)\\
& + \Ss_{\rm Ba2}(y\tS{t+1}|x\tS{t+1}; \pi\tSb{t}) \\
& - \DKL \Big[ \pi\tSb{t} || p_c \pi\tSb{0} + (1-p_c)\pi\tSb{t} \Big]  \\
& + C \big[ \pi\tSb{t} \big] - C \big[ \pi_\flatp \big],
\end{aligned} \end{equation} \end{linenomath*}
and our 2nd definition can be written as
\begin{linenomath*} \begin{equation}\ \begin{aligned}
\label{eq:prop_CC2}
\Ss_{\rm CC2}(y\tS{t+1}|&x\tS{t+1}; \pi\tSb{t}) = \\
&\Delta\Ss_{\rm Sh1}(y\tS{t+1}|x\tS{t+1}; \pi\tSb{t}) \\
& + \Ss_{\rm Ba2}(y\tS{t+1}|x\tS{t+1}; \pi\tSb{t}) \\
& - \DKL \Big[ \pi\tSb{t} || p_c \pi\tSb{0} + (1-p_c)\pi\tSb{t} \Big]\\
& + \DKL \Big[ \pi\tSb{t} || \pi\tSb{0} \Big].
\end{aligned} \end{equation} \end{linenomath*}
\end{myprop}

Proposition \ref{Prop:CC} conveys three important messages.
First, both definitions of the Confidence Corrected surprise depend on differences in the Shannon surprise as opposed to the Shannon surprise itself (c.f. first line in \autoref{eq:prop_CC1} and \autoref{eq:prop_CC2}).
Second, both definitions depend on the difference between the Bayesian surprise (i.e., the change in the belief given the new observation) and the \textit{a priori} expected change in the belief (because of the possibility of a change in the environment; c.f. second and third lines in \autoref{eq:prop_CC1} and \autoref{eq:prop_CC2}).
Third, both definitions regularize the contributions of Shannon surprise and Bayesian surprise by the relative confidence of the current belief compared to either the flat or the prior belief (c.f. the last line in \autoref{eq:prop_CC1} and \autoref{eq:prop_CC2}).
`Relative confidence' quantifies how different the current belief is with respect to a reference belief;
note that $C \big[ \pi\tSb{t} \big] - C \big[ \pi_\flatp \big] = \DKL \big[ \pi\tSb{t} || \pi_\flatp \big] $.

Hence, the Confidence Corrected surprise should be distinguishable from both the Shannon and the Bayesian surprise (for $p_c < 1$).
An interesting consequence of Proposition \ref{Prop:CC}, however, is that $\Ss_{\rm CC2}$ is identical to $\Ss_{\rm Ba2}$ when the environment becomes so volatile that its parameter changes at each time step (i.e., in the limit of $p_c \to 1$):
\begin{mycorrol}
\label{Corr:CC_Ba_pc1}
For the generative model of Definition \ref{def:gen_model}, when $p_c \to 1$, we have $\Ss_{\rm CC2}(y\tS{t+1}|x\tS{t+1}; \pi\tSb{t}) = \Ss_{\rm Ba2}(y\tS{t+1}| \allowbreak x\tS{t+1}; \pi\tSb{t})$.
\end{mycorrol}

\subsection{Minimized free energy}

Although an agent can perform computations over the joint probability distribution in \autoref{eq:gen_mod_1} and \autoref{eq:gen_mod_2}, finding the belief $\pi\tSb{t+1}(\theta)$ (i.e., the posterior distribution in \autoref{eq:belief}) can be computationally intractable \parencite{barber2012bayesian, liakoni2019approximate}.
Therefore, it has been argued that the brain uses approximate inference (instead of exact Bayesian inference) for finding the belief \parencite{mathys2011bayesian, liakoni2019approximate, friston2017active, daw2008pigeon, friston2010free, fiser2010statistically, faraji2018balancing, findling2019imprecise}.
An approximation of the belief $\pi\tSb{t+1}(\theta)$ can for example be found via variational inference \parencite{blei2017variational, mackay2003information} over a family of distributions $q(\theta ; \phi)$ parameterized by $\phi$.
Such approaches are popular in neuroscience studies of learning and inference in the brain \parencite{friston2010free, friston2017active, gershman2019does}.

Formally, in variational inference, the belief $\pi\tSb{t+1}(\theta)$ is approximated by $\hat{\pi}\tSb{t+1}(\theta) \coloneqq q(\theta ; \allowbreak \phi\tSb{t+1})$, where $\phi\tSb{t+1}$ is the minimizer of the variational loss or free energy, i.e., $\phi\tSb{t+1} \coloneqq \arg \min_\phi F\tSb{t+1}(\phi)$ \parencite{mackay2003information}.
To define $F\tSb{t+1}(\phi)$, we introduce a new notation:
 \begin{linenomath*} \begin{equation}\ \begin{aligned}
\PP_{\Theta\tS{t+1}}&\big(\theta, y\tS{t+1} | x\tS{t+1}; \pi \big) \coloneqq\\
&P_{Y|X}(y\tS{t+1}|x\tS{t+1}; \theta) \Big( p_c \pi\tSb{0}(\theta) + (1-p_c) \pi(\theta) \Big),
\end{aligned} \end{equation} \end{linenomath*}
where $\pi$ is an arbitrary distribution over the parameter space.
Using this notation, we can write the joint distribution over the observation and the parameter $\PP\tSb{t}\big(\theta\tS{t+1}, y\tS{t+1} | \allowbreak x\tS{t+1}\big)$ as
$\PP_{\Theta\tS{t+1}}\big(\theta\tS{t+1}, y\tS{t+1} | x\tS{t+1}; \pi\tSb{t} \big)$ and the updated belief $\pi\tSb{t+1}(\theta)$ as
$\PP_{\Theta\tS{t+1}}\big(\theta |  y\tS{t+1}, x\tS{t+1}; \pi\tSb{t} \big)$.
The variational loss or free energy can then be defined as \parencite{liakoni2019approximate, markovic2021empirical, sajid2021active}
 \begin{linenomath*} \begin{equation}\ \begin{aligned}
\label{eq:FE_loss}
F\tSb{t+1}(\phi) \coloneqq \EE_{q(.;\phi)} \Big[
&\log q(\Theta;\phi) -\\
&\log \PP_{\Theta\tS{t+1}}\big(\Theta, y\tS{t+1} | x\tS{t+1}; \hat{\pi}\tSb{t} \big)
\Big].
\end{aligned} \end{equation} \end{linenomath*}
For any value of $\phi$, one can show that \parencite{blei2017variational, sajid2021active}
\begin{linenomath*} \begin{equation}\ \begin{aligned}
\label{Eq:min_FE}
F\tSb{t+1}(\phi)  = \,
&\Ss_{\rm Sh1}(y\tS{t+1}|x\tS{t+1}; \hat{\pi}\tSb{t}) +\\
&\DKL \Big[ q(.;\phi) || \PP_{\Theta\tS{t+1}}\big(. | y\tS{t+1}, x\tS{t+1}; \hat{\pi}\tSb{t} \big) \Big]
\\
\geq \,
&\Ss_{\rm Sh1}(y\tS{t+1}|x\tS{t+1}; \hat{\pi}\tSb{t}),
\end{aligned} \end{equation} \end{linenomath*}
where the right side of the inequality is independent of $\phi$, and  $\PP_{\Theta\tS{t+1}}\big(. | y\tS{t+1}, x\tS{t+1}; \hat{\pi}\tSb{t} \big)$ is the exact Bayesian update of the belief (according to the generative model in Definition \ref{def:gen_model}) given the latest approximation of the belief $\hat{\pi}\tSb{t}$ \parencite{liakoni2019approximate, markovic2021empirical}.

The minimized free energy $F^* \coloneqq \min_\phi F\tSb{t+1}(\phi)$ has been interpreted as a measure of surprise \parencite{friston2010free, schwartenbeck2013exploration,friston2017active}, which, according to \autoref{Eq:min_FE}, can be seen as an approximation of $\Ss_{\rm Sh1}(y\tS{t+1}|x\tS{t+1}; \hat{\pi}\tSb{t})$.
The parametric family of $q(.;\phi)$ and its relation to the exact belief $\pi\tSb{t+1}$ determine how well $F^*$ approximates $\Ss_{\rm Sh1}(y\tS{t+1}|x\tS{t+1}; \hat{\pi}\tSb{t})$ (\autoref{fig:surprise_relation_schematic}B).
More precisely, the minimized free energy measures both how unlikely the new observation is (i.e., how large $\Ss_{\rm Sh1}(y\tS{t+1}|x\tS{t+1}; \allowbreak \hat{\pi}\tSb{t})$ is) and how imprecise the best parametric approximation of the belief $\hat{\pi}\tSb{t+1}$ is (i.e., how large $\DKL [ \hat{\pi}\tSb{t+1} || \allowbreak \PP_{\Theta\tS{t+1}}\big(. | y\tS{t+1}, x\tS{t+1}; \hat{\pi}\tSb{t} \big) ]$ is).
Therefore, the minimized free energy is in the category of belief-mismatch surprise measures (\autoref{fig:surprise_belief}).

\section{Taxonomy of surprise definitions}
\label{sec:conc_class}

\begin{figure*}[!t]
    \centering
    \includegraphics[width=1\textwidth]{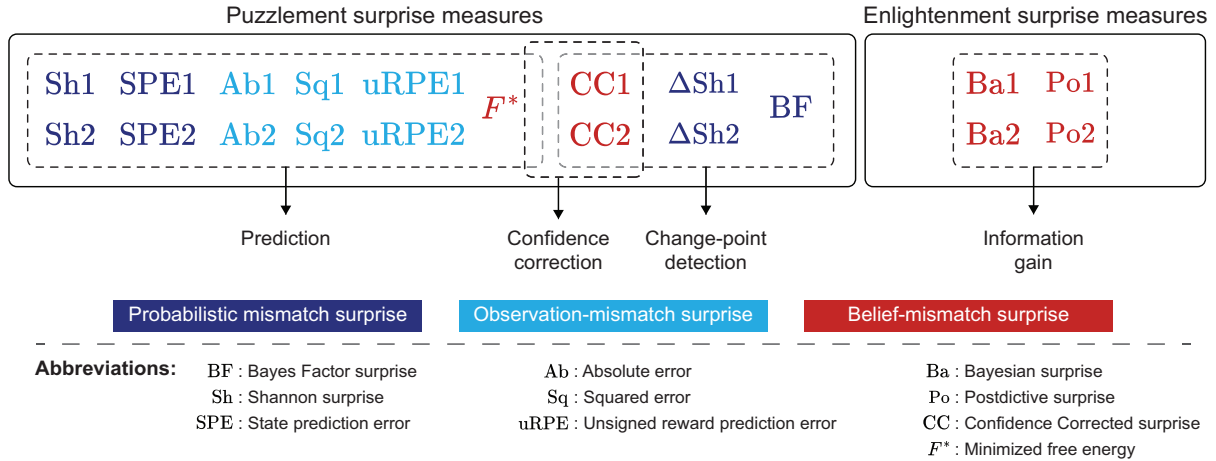}
    \caption{\textbf{Taxonomy of surprise definitions.}
    Measures of puzzlement surprise \parencite{faraji2018balancing} can be further classified into 3 sub-categories of surprise measures highlighting (i) prediction, (ii) change-point detection, and (iii) confidence correction.
    According to surprise measures focused on prediction, the agent's puzzle is finding the most accurate prediction of the next observation.
    According to surprise measures focused on change-point detection, the agent's puzzle is to detect environmental changes.
    Surprise measures focused on confidence correction do not determine a specific puzzle (change-point detection or accurate prediction, visualized by overlapping boxes) for the agent but stress that confidence should explicitly influence puzzlement.
    The enlightenment surprise measures can be seen as measures of information gain.
    In addition to the 18 definitions of surprise discussed in \autoref{sec:surprise_theory}, we included in the figure the difference in Shannon surprise ($\Delta$Sh1 and $\Delta$Sh2) introduced in Proposition \ref{Prop:dSh_BF}.
    Color code shows the technical classification presented in \autoref{fig:surprise_belief}.
\textit{Color should be used in print.}
    }
    \label{fig:surprise_puzzle}
\end{figure*}

In a unified framework, we discussed 10 previously proposed measures of surprise:
(1) the Bayes Factor surprise; (2) the Shannon surprise; (3) the State Prediction Error; (4) the Absolute and (5) the Squared error surprise; (6) the unsigned Reward Prediction Error; (7) the Bayesian surprise; (8) the Postdictive surprise; (9) the Confidence Corrected surprise; and (10) the Minimized Free Energy.
We considered different ways to define some of these measures in volatile environments and, overall, analyzed 18 different definitions of surprise.
In this section, we propose a taxonomy of these 18 definitions and classify them into four main categories regarding the semantic of what they quantify (\autoref{fig:surprise_puzzle}).

Measures of surprise in neuroscience have been previously divided into two categories \parencite{hurley2011inside, faraji2018balancing, gijsen2021}: `puzzlement' and `enlightenment' surprise.
Puzzlement surprise measures how puzzling a new observation is for an agent, whereas enlightenment surprise measures how much the new observation has enlightened the agent and changed its belief -- a concept closely linked but not identical to the `Aha! moment' \parencite{kounios2009aha, dubey2021aha}.
The Bayesian and the Postdictive surprise can be categorized as enlightenment surprise since both quantify information gain (\autoref{fig:surprise_puzzle}).
Based on our theoretical analyses, however, we suggest to further divide measures of puzzlement surprise into 3 sub-categories (\autoref{fig:surprise_puzzle}):

\textbf{i. `Prediction surprise'}
quantifies how unpredicted, unexpected, or unlikely the new observation is.
This category includes the Shannon surprise, State Prediction Error, the Minimized Free Energy, and all observation-mismatch surprise measures (\autoref{fig:surprise_puzzle}).
According to these measures, the agent's puzzle is to find the most accurate predictions of the next observations.
Surprise in natural language is defined as `the feeling or emotion excited by something unexpected' \parencite{oxford_surp}.
If we focus on the term `unexpected', identify it with `unlikely under the current belief', and neglect the terms `feeling' and `emotion', then the quality measured by prediction surprise is closely related to the definition of surprise in natural language.

\textbf{ii. `Change-point detection surprise'}
quantifies relative unlikeliness of the new observation and are designed to modulate the learning rate and to identify environmental changes.
This category includes the Bayes Factor surprise and the difference in Shannon surprise (c.f. Corollary \ref{Corr:gamma_Sh}; \autoref{fig:surprise_puzzle}).
According to these measures, the agent's puzzle is to detect environmental changes.

\textbf{iii. `Confidence correction surprise'}
explicitly accounts for the agent's confidence.
The idea is that higher confidence (or higher commitment to a belief) leads to more puzzlement, where the puzzle is either to detect environmental changes or to find the most accurate prediction.
\cite{faraji2018balancing} argue, using a thought experiment, that such an explicit account for confidence is crucial to explain our perception of surprise.
The only current candidates of this category are $\Ss_{\rm CC1}$ and $\Ss_{\rm CC2}$ that assume that the agent's puzzle is to detect environmental changes (c.f. Proposition \ref{Prop:CC});
but we anticipate that more examples in this category might be found in the future.

While our proposed taxonomy is solely conceptual and based on the theoretical properties of different definitions, we note that there have been a significant number of studies investigating the neural and physiological correlates of prediction \parencite{mars2008trial, kopp2013electrophysiological, kolossa2015computational, modirshanechi2019trial, gijsen2021, maheu2019brain, meyniel2020brain, mousavi2020brain, konovalov2018neurocomputational, loued2020information, glascher2010states}, change-point detection \parencite{nassar2012rational, xu2021novelty, liakoni2022brain}, confidence correction \parencite{gijsen2021}, and information gain \parencite{ostwald2012evidence, kolossa2015computational, gijsen2021, visalli2021EEG, nour2018dopaminergic, o2013dissociable} surprise measures (\autoref{fig:experiments}).
We, therefore, speculate that at least one measure from each of these categories is computed in the brain but potentially through different neural pathways and to be used for different brain functions.

\section{Discussion}
\label{sec:disc}
What does it formally mean to be surprised?
And how do existing definitions of surprise relate to each other?
To address these questions, we reviewed 18 definitions of surprise in a unifying mathematical framework and studied their similarities and differences.
We showed that several extensions of known surprise measures to volatile environments are possible and potentially relevant;
hence, further experimental evidence is needed to elucidate the relevance of precise definitions of surprise for brain research.
Based on how different definitions depend on the belief $\pi\tSb{t}$, we divided them into three groups of probabilistic mismatch, observation-mismatch, and belief-mismatch surprise measures (\autoref{fig:surprise_belief}).
We then showed how these measures relate to each other theoretically and, more importantly, under which conditions they are strictly increasing functions of each other (i.e., they become experimentally indistinguishable -- \autoref{fig:surprise_relation_schematic} and \autoref{tab:experiments}).
We further proposed a taxonomy of surprise definitions by a conceptual classification into four main categories (\autoref{fig:surprise_puzzle}):
(i) prediction surprise, (ii) change-point detection surprise, (iii) confidence-corrected surprise, and (iv) information gain surprise.

It is believed that surprise has important computational roles in different brain functions such as adaptive learning \parencite{iigaya2016adaptive, gerstner2018eligibility}, exploration \parencite{dubey2020understanding, gottlieb2018towards}, memory formation \parencite{rouhani2021signed}, and memory segmentation \parencite{antony2021behavioral}.
Our results propose a diverse toolkit and a refined terminology to theoreticians and computational scientist to model and discuss the different functions of surprise and their biological implementation.
For instance, it has been argued that the computation of observation-mismatch surprise measures is biologically more plausible than more abstract measures such as Shannon surprise \parencite{iigaya2016adaptive}.
Our results identify conditions under which observation-mismatch surprise measures behave identically to probabilistic mismatch surprise measures that are optimal for adaptive learning (c.f. \autoref{fig:surprise_relation_schematic}B, Proposition \ref{Prop:update}, and Corollary \ref{Corr:gamma_Sh});
such insights can be exploited in future network models of adaptive behavior.

Moreover, our results can be used to design novel theory-driven experiments where different measures of surprise make different predictions.
Importantly, most of the previous experimental studies have focused on one measure of surprise and its role and signatures in behavioral and physiological measurements.
The examples that considered more than one surprise measure \parencite{mars2008trial, ostwald2012evidence, kolossa2015computational, gijsen2021, mousavi2020brain} have mainly focused on model-selection methods to compare different models and did not look for \textit{fundamentally} different predictions of these measures -- see \cite{visalli2021EEG} for an exception.
Even if two surprise measures are formally distinguishable, it may be that, in a given experimental set-up, the number of samples or effect size are not big enough to extract the quantitative differences between the two.
For example, $\Ss_{\rm BF}$ and $\Ss_{\rm Sh1}$ are distinguishable for any prior marginal distributions other than uniform distribution (\autoref{fig:surprise_relation_schematic}B), but, in practice, the distinction is hard to detect for nearly-uniform priors.
Our theoretical framework enables us to go further and design experiments that enable to dissociate different surprise measures based on their \textit{qualitatively} different predictions and to avoid experiments where different measures are either formally or practically indistinguishable.

\section*{Acknowledgement}
AM is grateful to Vasiliki Liakoni, Martin Barry, and Valentin Schmutz for many useful discussions in the course of the last few years, and to Andrew Barto for insightful discussions during and after EPFL Neuro Symposium 2021 on ``Surprise, Curiosity and Reward: from Neuroscience to AI''.
This research was supported by Swiss National Science Foundation (no. $200020\_184615$).

\section*{Competing Interests statement}
The authors declare no competing interests.

\appendix
\section{Proofs}
\label{sec:app_proofs}
In this appendix, we provide proofs for our Propositions and Corollaries mentioned in the main text.
We also provide further results for the postdictive surprise in Lemma \ref{Lemma:Po_Sh}.

\subsection{Proof of Proposition \ref{Prop:update}}
The proof is in essence the same as the proof of Proposition 1 of \cite{liakoni2019approximate}.
We write
\begin{linenomath*} \begin{equation}\
\begin{aligned}
\label{eq:proofs_pit1_exp}
\pi\tSb{t+1}(\theta) &= \PP\tSb{t+1}(\Theta\tS{t+1}=\theta) \\
= &\PP\tSb{t+1}(\Theta\tS{t+1}=\theta | C\tS{t+1} = 0) \PP\tSb{t+1}( C\tS{t+1} = 0) +\\
& \PP\tSb{t+1}(\Theta\tS{t+1}=\theta | C\tS{t+1} = 1) \PP\tSb{t+1}( C\tS{t+1} = 1).
\end{aligned}
\end{equation} \end{linenomath*}
We use Bayes' rule and write $\PP\tSb{t+1}(\Theta\tS{t+1}=\theta | C\tS{t+1} = 0)$ (c.f. the 1st term in \autoref{eq:proofs_pit1_exp}) as
\begin{linenomath*} \begin{equation}\
\begin{aligned}
\PP\tSb{t+1}&(\Theta\tS{t+1}=\theta | C\tS{t+1} = 0) \\
= & \PP\tSb{t}(\Theta\tS{t+1}=\theta | C\tS{t+1} = 0, x\tS{t+1}, y\tS{t+1})\\
= & \frac{ \PP\tSb{t}( y\tS{t+1}|C\tS{t+1} = 0, x\tS{t+1}, \Theta\tS{t+1}=\theta )}
{\PP\tSb{t}( y\tS{t+1}|C\tS{t+1} = 0, x\tS{t+1})} \times \\
&\PP\tSb{t}(\Theta\tS{t+1}=\theta | C\tS{t+1} = 0, x\tS{t+1})\\
= & \frac{P_{Y|X}(y\tS{t+1}|x\tS{t+1};\theta) \pi\tSb{t}(\theta)}{P(y\tS{t+1}|x\tS{t+1};\pi\tSb{t})} = \pi\tSb{t+1}_{\rm integration}(\theta),
\end{aligned}
\end{equation} \end{linenomath*}
and similarly
\begin{linenomath*} \begin{equation}\
\begin{aligned}
\PP\tSb{t+1}(\Theta\tS{t+1}=\theta | C\tS{t+1} = 1) &= \frac{P_{Y|X}(y\tS{t+1}|x\tS{t+1};\theta) \pi\tSb{0}(\theta)}{P(y\tS{t+1}|x\tS{t+1};\pi\tSb{0})}\\
&= \pi\tSb{t+1}_{\rm reset}(\theta).
\end{aligned}
\end{equation} \end{linenomath*}
Then, for $\PP\tSb{t+1}( C\tS{t+1} = 1)$ and $\PP\tSb{t+1}( C\tS{t+1} = 0) = 1 - \PP\tSb{t+1}( C\tS{t+1} = 1)$ we have
\begin{linenomath*} \begin{equation}\
\begin{aligned}
&\PP\tSb{t+1}( C\tS{t+1} = 1)\\
& = \PP\tSb{t}(C\tS{t+1} = 1| x\tS{t+1}, y\tS{t+1}) \\
& = \frac{p_c P(y\tS{t+1}|x\tS{t+1};\pi\tSb{0})}{(1-p_c) P(y\tS{t+1}|x\tS{t+1};\pi\tSb{t})+ p_c P(y\tS{t+1}|x\tS{t+1};\pi\tSb{0})} \\
&= \frac{m \Ss_{\rm BF}(y\tS{t+1}|x\tS{t+1}; \pi\tSb{t})}{1 + m\Ss_{\rm BF}(y\tS{t+1}|x\tS{t+1}; \pi\tSb{t})} = \gamma\tS{t+1}
\end{aligned}
\end{equation} \end{linenomath*}
with $m = \frac{p_c}{1-p_c}$.
Therefore, the proof is complete by substituting these terms in \autoref{eq:proofs_pit1_exp}.
$\hfill \blacksquare$

\subsection{Proof of Proposition \ref{Prop:dSh_BF}}
Based on the definition of the adaptation rate $\gamma\tS{t+1}$ (c.f. Proposition \ref{Prop:update}), we have
\begin{linenomath*} \begin{equation}\
\begin{aligned}
\Ss_{\rm BF}(y\tS{t+1}|x\tS{t+1}; \pi\tSb{t}) &= \frac{1-p_c}{p_c} \frac{\gamma\tS{t+1}}{1-\gamma\tS{t+1}}.
\end{aligned}
\end{equation} \end{linenomath*}
For the difference in the 1st definition of the Shannon surprise (c.f. \autoref{eq:Sh1}), we can write
\begin{linenomath*} \begin{equation}\
\begin{aligned}
\label{eq:proof_2_Sh1}
&\Delta \Ss_{\rm Sh1}(y\tS{t+1}|x\tS{t+1}; \pi\tSb{t}) \\
&=\Ss_{\rm Sh1}(y\tS{t+1}|x\tS{t+1}; \pi\tSb{t}) - \Ss_{\rm Sh1}(y\tS{t+1}|x\tS{t+1}; \pi\tSb{0}) \\
&=\log \Big( \frac
{P(y\tS{t+1}|x\tS{t+1};\pi\tSb{0})}
{p_c P(y\tS{t+1}|x\tS{t+1};\pi\tSb{0}) + (1-p_c) P(y\tS{t+1}|x\tS{t+1};\pi\tSb{t})}
\Big) \\
&=\log \frac{\gamma\tS{t+1}}{p_c}.
\end{aligned}
\end{equation} \end{linenomath*}
As a result, we have $\gamma\tS{t+1} = p_c \exp \Delta \Ss_{\rm Sh1}(y\tS{t+1}|x\tS{t+1}; \pi\tSb{t})$ and hence
\begin{linenomath*} \begin{equation}\
\begin{aligned}
\Ss_{\rm BF}(y\tS{t+1}|&x\tS{t+1}; \pi\tSb{t}) \\
&= \frac{(1-p_c)\exp \Delta \Ss_{\rm Sh1}(y\tS{t+1}|x\tS{t+1}; \pi\tSb{t})}{1-p_c\exp \Delta \Ss_{\rm Sh1}(y\tS{t+1}|x\tS{t+1}; \pi\tSb{t})}.
\end{aligned}
\end{equation} \end{linenomath*}
The proof is more straightforward for the difference in the 2nd definition (c.f.  \autoref{eq:Sh2}) where we have
\begin{linenomath*} \begin{equation}\
\begin{aligned}
\label{eq:proof_2_Sh2}
\Delta &\Ss_{\rm Sh2}(y\tS{t+1}|x\tS{t+1}; \pi\tSb{t}) \\
& = \Ss_{\rm Sh2}(y\tS{t+1}|x\tS{t+1}; \pi\tSb{t}) - \Ss_{\rm Sh2}(y\tS{t+1}|x\tS{t+1}; \pi\tSb{0})\\
& = \log \Big( \frac
{P(y\tS{t+1}|x\tS{t+1};\pi\tSb{0})}
{P(y\tS{t+1}|x\tS{t+1};\pi\tSb{t})}
\Big) = \log \Ss_{\rm BF}(y\tS{t+1}|x\tS{t+1}; \pi\tSb{t}).
\end{aligned}
\end{equation} \end{linenomath*}
Therefore, the proof is complete.
$\hfill \blacksquare$

\subsection{Proof of Proposition \ref{Prop:SPE_Sh}}

Based on the definitions of the two versions of the Shannon surprise (c.f. \autoref{eq:Sh1} and  \autoref{eq:Sh2}), we have
\begin{linenomath*} \begin{equation}\
\begin{aligned}
\PP\tSb{t}\big(y\tS{t+1}|x\tS{t+1} \big) &= \exp \Big( - \Ss_{\rm Sh1}(y\tS{t+1}|x\tS{t+1}; \pi\tSb{t}) \Big),\\
P(y\tS{t+1}|x\tS{t+1};\pi\tSb{t}) &= \exp \Big( - \Ss_{\rm Sh2}(y\tS{t+1}|x\tS{t+1}; \pi\tSb{t}) \Big).
\end{aligned}
\end{equation} \end{linenomath*}
The proof is complete by using these equations and replacing the probabilities in \autoref{eq:SPE1} and  \autoref{eq:SPE2}.
$\hfill \blacksquare$

\subsection{Proof of Proposition \ref{Prop:Ab_Sq_Cat}}
For a categorical task with $N$ categories and one-hot coded observations, we have (c.f. \autoref{eq:E1} and \autoref{eq:E2})
\begin{linenomath*} \begin{equation}\
\begin{aligned}
E_1 [Y\tS{t+1}] = \Big[ &p_c P(n|x\tS{t+1};\pi\tSb{0}) + \\
                        &(1-p_c) P(n|x\tS{t+1};\pi\tSb{t}) \Big]_{n=1}^N\\
E_2 [Y\tS{t+1}] = \Big[ &P(n|x\tS{t+1};\pi\tSb{t}) \Big]_{n=1}^N\\
\end{aligned}
\end{equation} \end{linenomath*}
where $z = [z_n]_{n=1}^N$ is an $N$-dimensional vector with $z_n$ the $n$th element.
To be able to prove the proposition for $E_1 [Y\tS{t+1}]$ and $E_2 [Y\tS{t+1}]$ simultaneously, we define $E_i [Y\tS{t+1}] = [p_{i,n}]_{n=1}^N$, where $p_{1,n} =  p_c P(n|x\tS{t+1};\pi\tSb{0}) + (1-p_c) \allowbreak P(n|x\tS{t+1};\allowbreak\pi\tSb{t})$ and $p_{2,n} =  P(n|x\tS{t+1};\pi\tSb{t})$.

We show the one-hot coded vector corresponding to category $m \in \{1,...,N\}$ by $e_m$.
For the absolute error surprise, we have (c.f. \autoref{eq:Ab_Sq})
\begin{linenomath*} \begin{equation}\
\begin{aligned}
\Ss_{{\rm Ab}i}&(y\tS{t+1}=e_m|x\tS{t+1}; \pi\tSb{t}) = \sum_{n=1}^N |\delta_{m,n} - p_{i,n}|\\
&= |1 - p_{i,m}| + \sum_{n=1, n \neq m}^N p_{i,n}\\
&= 2(1 - p_{i,m}),
\end{aligned}
\end{equation} \end{linenomath*}
which is the same as $2 \Ss_{{\rm SPE}i}(y\tS{t+1}=e_m|x\tS{t+1}; \pi\tSb{t})$ (c.f. \autoref{eq:SPE1} and \autoref{eq:SPE2}).

For the squared error surprise, we have (c.f. \autoref{eq:Ab_Sq})
\begin{linenomath*} \begin{equation}\
\begin{aligned}
\Ss_{{\rm Sq}i}(y\tS{t+1}=e_m|&x\tS{t+1}; \pi\tSb{t}) = \sum_{n=1}^N (\delta_{m,n} - p_{i,n})^2 \\
& = (1 - p_{i,m})^2 + \sum_{n=1, n \neq m}^N p_{i,n}^2 \\
& = 2(1 - p_{i,m}) + || [p_{i,n}]_{n=1}^N ||_2^2 - 1,
\end{aligned}
\end{equation} \end{linenomath*}
where we have $2(1 - p_{i,m})  = 2 \Ss_{{\rm SPE}i}(y\tS{t+1}=e_m|x\tS{t+1}; \pi\tSb{t})$ and
\begin{linenomath*} \begin{equation}\
\begin{aligned}
{\rm Conf.}\Big[P(.|x\tS{t+1};\pi\tSb{t}) \Big] = || [p_{i,n}]_{n=1}^N ||_2^2 - 1
\end{aligned}
\end{equation} \end{linenomath*}
shows the $\ell_2$-norm of the estimate vector $[p_{i,n}]_{n=1}^N$ as a measure of confidence;
$|| [p_{i,n}]_{n=1}^N ||_2^2$ takes its maximum value when the prediction has a probability of 1 for one category and zero for the rest and takes its minimum when it is distributed uniformly over all categories.
Therefore, the proof is complete.
$\hfill \blacksquare$

\subsection{Proof of Proposition \ref{Prop:Sq_Gau}}
Assume that $Y\tS{t+1} \in \RR^N$, given the cue $x\tS{t+1}$ and the belief $\pi\tSb{t}$, has a Gaussian distribution with a covariance matrix $\sigma^2 I$, i.e.,
\begin{linenomath*} \begin{equation}\
\begin{aligned}
P(y\tS{t+1}| x\tS{t+1}; \pi\tSb{t}) = \Nn\Big(y\tS{t+1}; E_2 [Y\tS{t+1}] , \sigma I \Big).
\end{aligned}
\end{equation} \end{linenomath*}
We then have
\begin{linenomath*} \begin{equation}\
\begin{aligned}
\Ss_{\rm Sh2}(y\tS{t+1}|&x\tS{t+1}; \pi\tSb{t}) =
- \log \Nn\Big( y\tS{t+1}; E_2 [Y\tS{t+1}] , \sigma I \Big)\\
&= \frac{N}{2} \log \big(2\pi\sigma \big)+ \frac{|| y\tS{t+1} - E_2 [Y\tS{t+1}] ||_2^2}{2\sigma^2}\\
&=
a + b \Ss_{{\rm Sq},2}(y\tS{t+1}=e_m|x\tS{t+1}; \pi\tSb{t}),
\end{aligned}
\end{equation} \end{linenomath*}
where $a= N \log \big(2\pi\sigma \big)/2$ and $b = 1/(2\sigma^2)$.
Therefore, the proof is complete.
$\hfill \blacksquare$

\subsection{Proof of Proposition \ref{Prop:Ab_Sq}}
Using the definition of the two surprise measures in \autoref{eq:Ab_Sq}, we have, for $y\tS{t+1} \in \RR$,
\begin{linenomath*} \begin{equation}\
\begin{aligned}
\Ss_{{\rm Sq}i}&(y\tS{t+1}|x\tS{t+1}; \pi\tSb{t}) = ||y\tS{t+1} -  E_i [Y\tS{t+1}]||^2_2\\
&= |y\tS{t+1} -  E_i [Y\tS{t+1}]|^2 =
\Ss_{{\rm Ab}i}(y\tS{t+1}|x\tS{t+1}; \pi\tSb{t})^2.
\end{aligned}
\end{equation} \end{linenomath*}
Therefore, the proof is complete.
$\hfill \blacksquare$

\subsection{Proof of Proposition \ref{Prop:uRPE}}
Using the definition of the uRPE and the absolute error surprise in \autoref{eq:Ab_Sq} and \autoref{eq:uRPE_comp}, we have
\begin{linenomath*} \begin{equation}\
\begin{aligned}
\Ss&_{{\rm Ab}i}(y\tS{t+1}|x\tS{t+1}; \pi\tSb{t}) = ||y\tS{t+1} -  E_i [Y\tS{t+1}]||_1\\
&= |\Tilde{r}\tS{t+1} -  E_i [\Tilde{R}\tS{t+1}]| + ||s\tS{t+1} -  E_i [S\tS{t+1}]||_1\\
&= \Ss_{{\rm uRPE}i}(y\tS{t+1}|x\tS{t+1}; \pi\tSb{t}) + \Ss_{{\rm Ab}i}(s\tS{t+1}|x\tS{t+1}; \pi\tSb{t}),
\end{aligned}
\end{equation} \end{linenomath*}
which complete the proof for the absolute error surprise.
Then, we can similarly write
\begin{linenomath*} \begin{equation}\
\begin{aligned}
\Ss&_{{\rm Sq}i}(y\tS{t+1}|x\tS{t+1}; \pi\tSb{t}) = ||y\tS{t+1} -  E_i [Y\tS{t+1}]||^2_2\\
&= |\Tilde{r}\tS{t+1} -  E_i [\Tilde{R}\tS{t+1}]|^2 + ||s\tS{t+1} -  E_i [S\tS{t+1}]||^2_2\\
&= \Ss_{{\rm uRPE}i}(y\tS{t+1}|x\tS{t+1}; \pi\tSb{t})^2 + \Ss_{{\rm Sq}i}(s\tS{t+1}|x\tS{t+1}; \pi\tSb{t}).
\end{aligned}
\end{equation} \end{linenomath*}
Therefore, the proof is complete.
$\hfill \blacksquare$

\subsection{Proof of Proposition \ref{Prop:Ba_Sh}}

For the 1st definition of the Bayesian surprise (c.f. \autoref{eq:Ba1}), we have
\begin{linenomath*} \begin{equation}\
\begin{aligned}
\Ss_{\rm Ba1}(y\tS{t+1}|x\tS{t+1}; \pi\tSb{t}) &= \DKL \Big[ \PP\tSb{t}_{\Theta\tS{t+1}} || \PP\tSb{t+1}_{\Theta\tS{t+1}} \Big]\\
&= \EE_{\PP\tSb{t}} \Big[  \log \frac{\PP\tSb{t}\big( \Theta\tS{t+1} \big)}{\PP\tSb{t+1}\big( \Theta\tS{t+1} \big)} \Big].
\end{aligned}
\end{equation} \end{linenomath*}
We know
\begin{linenomath*} \begin{equation}\
\begin{aligned}
\PP\tSb{t}_{\Theta\tS{t+1}} = p_c \pi\tSb{0}+ (1-p_c) \pi\tSb{t},
\end{aligned}
\end{equation} \end{linenomath*}
and
\begin{linenomath*} \begin{equation}\
\begin{aligned}
\label{eq:proof_4_prob_ratio}
\PP\tSb{t+1}\big( \theta\tS{t+1}  \big) &=
\frac
{\PP\tSb{t}\big( \theta\tS{t+1} \big) P_{Y|X}\big( y\tS{t+1} | x\tS{t+1}; \theta\tS{t+1}\big)}
{ \PP\tSb{t}\big( y\tS{t+1} | x\tS{t+1}\big)}\\
& \Rightarrow \\
\frac{\PP\tSb{t+1}\big( \theta\tS{t+1}  \big)}{\PP\tSb{t}\big( \theta\tS{t+1} \big)} &=
\frac{P_{Y|X}\big( y\tS{t+1} | x\tS{t+1}; \theta\tS{t+1}\big)}
{ \PP\tSb{t}\big( y\tS{t+1} | x\tS{t+1}\big)}.
\end{aligned}
\end{equation} \end{linenomath*}
We, therefore, have
\begin{linenomath*} \begin{equation}\
\begin{aligned}
\Ss_{\rm Ba1}(y\tS{t+1}&|x\tS{t+1}; \pi\tSb{t}) =
- p_c \EE_{\pi\tSb{0}}  \Big[\log P_{Y|X}(y\tS{t+1}|x\tS{t+1};\Theta) \Big] \\
&- (1-p_c) \EE_{\pi\tSb{t}}  \Big[\log P_{Y|X}(y\tS{t+1}|x\tS{t+1};\Theta) \Big]\\
&+ \log \PP\tSb{t}\big( y\tS{t+1} | x\tS{t+1}\big),
\end{aligned}
\end{equation} \end{linenomath*}
which is equivalent to (c.f. \autoref{eq:Sh1} and  \autoref{eq:Sh2})
\begin{linenomath*} \begin{equation}\
\begin{aligned}
\Ss_{\rm Ba1}(y\tS{t+1}|x\tS{t+1};& \pi\tSb{t}) =
p_c \EE_{\pi\tSb{0}} \Big[ \Ss_{\rm Sh2}(y\tS{t+1}|x\tS{t+1}; \delta_{\{\Theta\}}) \Big] \\
&+ (1-p_c) \EE_{\pi\tSb{t}} \Big[ \Ss_{\rm Sh2}(y\tS{t+1}|x\tS{t+1}; \delta_{\{\Theta\}}) \Big]\\
&- \Ss_{\rm Sh1}(y\tS{t+1}|x\tS{t+1}; \pi\tSb{t}).
\end{aligned}
\end{equation} \end{linenomath*}
For the 2nd definition of the Bayesian surprise (c.f. \autoref{eq:Ba2}), we have
\begin{linenomath*} \begin{equation}\
\begin{aligned}
\Ss_{\rm Ba2}(y\tS{t+1}|x\tS{t+1}; \pi\tSb{t}) &= \DKL \Big[ \pi\tSb{t} || \pi\tSb{t+1}  \Big]\\
&=
\EE_{\pi\tSb{t}} \Big[ \log \frac{\pi\tSb{t}\big( \Theta \big)}{\pi\tSb{t+1}\big( \Theta \big)} \Big].
\end{aligned}
\end{equation} \end{linenomath*}
We use \autoref{eq:pi_t_weighted} and \autoref{eq:proof_4_prob_ratio} and write
\begin{linenomath*} \begin{equation}\
\begin{aligned}
\Ss_{\rm Ba2}(y\tS{t+1}&|x\tS{t+1}; \pi\tSb{t}) =
- \EE_{\pi\tSb{t}} \Big[\log P_{Y|X}(y\tS{t+1}|x\tS{t+1};\Theta)\Big] \\
&+ \log \PP\tSb{t}\big( y\tS{t+1} | x\tS{t+1}\big) \\
&+ \EE_{\pi\tSb{t}} \Big[\log \frac{\pi\tSb{t}\big( \Theta \big)}{p_c \pi\tSb{0}\big( \Theta \big) + (1-p_c) \pi\tSb{t}\big( \Theta \big)} \Big],
\end{aligned}
\end{equation} \end{linenomath*}
which is equivalent to (c.f. \autoref{eq:Sh1} and  \autoref{eq:Sh2})
\begin{linenomath*} \begin{equation}\
\begin{aligned}
\label{eq:proof_4_Ba2_final}
\Ss_{\rm Ba2}(y\tS{t+1}|&x\tS{t+1}; \pi\tSb{t}) =  \EE_{\pi\tSb{t}} \Big[ \Ss_{\rm Sh2}(y\tS{t+1}|x\tS{t+1}; \delta_{\{\Theta\}}) \Big]\\
&- \Ss_{\rm Sh1}(y\tS{t+1}|x\tS{t+1}; \pi\tSb{t})\\
&+ \DKL \Big[ \pi\tSb{t} || p_c \pi\tSb{0} + (1-p_c)\pi\tSb{t} \Big].
\end{aligned}
\end{equation} \end{linenomath*}
Therefore, the proof is complete.
$\hfill \blacksquare$

\subsection{Proof of Proposition \ref{Prop:CC}}
First, we prove the statement for the 2nd definition of the Confidence Corrected surprise (c.f. \autoref{eq:CC2}) for which we have
\begin{linenomath*} \begin{equation}\
\begin{aligned}
\Ss_{\rm CC2}(y\tS{t+1}|x\tS{t+1}; \pi\tSb{t}) &= \DKL \Big[ \pi\tSb{t} || \pi\tSb{t+1}_{\rm reset} \Big] \\
&=
\EE_{\pi\tSb{t}} \Big[ \log \frac{\pi\tSb{t}\big( \Theta \big)}{\pi\tSb{t+1}_{\rm reset}\big( \Theta \big)} \Big].
\end{aligned}
\end{equation} \end{linenomath*}
Using the definition of $\pi\tSb{t+1}_{\rm reset}$ in Proposition \ref{Prop:update}, we can write
\begin{linenomath*} \begin{equation}\
\begin{aligned}
\Ss_{\rm CC2}(y\tS{t+1}|x\tS{t+1}; \pi\tSb{t}) =
&- \EE_{\pi\tSb{t}} \Big[ \log P_{Y|X}(y\tS{t+1}|x\tS{t+1};\Theta) \Big]\\
&+ \log P\big( y\tS{t+1} | x\tS{t+1}; \pi\tSb{0} \big)\\
&+ \EE_{\pi\tSb{t}} \Big[ \log \frac{\pi\tSb{t}\big( \Theta \big)}{\pi\tSb{0}\big( \Theta \big)} \Big],
\end{aligned}
\end{equation} \end{linenomath*}
which is equivalent to (c.f. \autoref{eq:Sh1} and  \autoref{eq:Sh2})
\begin{linenomath*} \begin{equation}\
\begin{aligned}
\Ss_{\rm CC2}(y\tS{t+1}|x\tS{t+1}; \pi\tSb{t}) = & \EE_{\pi\tSb{t}} \Big[ \Ss_{\rm Sh2}(y\tS{t+1}|x\tS{t+1}; \delta_{\{\Theta\}}) \Big] \\
&- \Ss_{\rm Sh1}(y\tS{t+1}|x\tS{t+1}; \pi\tSb{0})\\
&+ \DKL \Big[ \pi\tSb{t} ||  \pi\tSb{0} \Big].
\end{aligned}
\end{equation} \end{linenomath*}
Now, we can replace $\EE_{\pi\tSb{t}} \Big[ \Ss_{\rm Sh2}(y\tS{t+1}|x\tS{t+1}; \delta_{\{\Theta\}}) \Big]$ by using \autoref{eq:proof_4_Ba2_final} and have
\begin{linenomath*} \begin{equation}\
\begin{aligned}
\Ss_{\rm CC2}(y\tS{t+1}|x\tS{t+1}; \pi\tSb{t}) = & \, \Ss_{\rm Sh1}(y\tS{t+1}|x\tS{t+1}; \pi\tSb{t})\\
&- \Ss_{\rm Sh1}(y\tS{t+1}|x\tS{t+1}; \pi\tSb{0})\\
&+\Ss_{\rm Ba2}(y\tS{t+1}|x\tS{t+1}; \pi\tSb{t}) \\
&- \DKL \Big[ \pi\tSb{t} || p_c \pi\tSb{0} + (1-p_c)\pi\tSb{t} \Big]\\
&+ \DKL \Big[ \pi\tSb{t} || \pi\tSb{0} \Big],
\end{aligned}
\end{equation} \end{linenomath*}
which is the same as \autoref{eq:prop_CC2}.
For the 1st definition of the Confidence Corrected surprise (c.f. \autoref{eq:CC1}), we can repeat all steps to have
\begin{linenomath*} \begin{equation}\
\begin{aligned}
\Ss_{\rm CC1}(y\tS{t+1}|x\tS{t+1}; \pi\tSb{t}) = & \, \Ss_{\rm Sh1}(y\tS{t+1}|x\tS{t+1}; \pi\tSb{t})\\
&- \Ss_{\rm Sh1}(y\tS{t+1}|x\tS{t+1}; \pi_\flatp)\\
&+\Ss_{\rm Ba2}(y\tS{t+1}|x\tS{t+1}; \pi\tSb{t}) \\
&- \DKL \Big[ \pi\tSb{t} || p_c \pi\tSb{0} + (1-p_c)\pi\tSb{t} \Big]\\
&+\DKL \Big[ \pi\tSb{t} || \pi_\flatp \Big].
\end{aligned}
\end{equation} \end{linenomath*}
If $\pi\tSb{t}$ is absolutely continuous with respect to $\pi_\flatp$, then we have $\DKL \Big[ \pi\tSb{t} || \pi_\flatp \Big] = C\Big[ \pi\tSb{t} \Big] - C\Big[ \pi_\flatp \Big]$, which completes the proof.
$\hfill \blacksquare$

\subsection{Proof of Corollary \ref{Corr:gamma_Sh}}
The corollary is the direct conclusion of \autoref{eq:proof_2_Sh1} and \autoref{eq:proof_2_Sh2}.
$\hfill \blacksquare$

\subsection{Proof of Corollary \ref{Corr:flat_prior}}
Let us show the set of possible observations by $\Yy$.
We assume that $\Yy$ is bounded, i.e., $|\Yy|<\infty$.
By assumption, we have $P(y\tS{t+1}|x\tS{t+1};\pi\tSb{0}) = 1/|\Yy|$.
We therefore (using \autoref{eq:BF}, \autoref{eq:Sh1}, and \autoref{eq:Sh2}) have
\begin{linenomath*} \begin{equation}\
\begin{aligned}
\Ss_{\rm Sh1}(y\tS{t+1}&|x\tS{t+1}; \pi\tSb{t})\\
&= \log \frac{m \Ss_{\rm BF}(y\tS{t+1}|x\tS{t+1}; \pi\tSb{t})}{1 + m \Ss_{\rm BF}(y\tS{t+1}|x\tS{t+1}; \pi\tSb{t})} + \log \frac{|\Yy|}{p_c},\\
\Ss_{\rm Sh2}(y\tS{t+1}&|x\tS{t+1}; \pi\tSb{t}) = \log \Ss_{\rm BF}(y\tS{t+1}|x\tS{t+1}; \pi\tSb{t}) + \log |\Yy|.
\end{aligned}
\end{equation} \end{linenomath*}
Both mappings are strictly increasing.
Therefore, the proof is complete.
$\hfill \blacksquare$

\subsection{Proof of Corollary \ref{Corr:CC_Ba_pc1}}
In the limit of $p_c \to 1$, we have $\Ss_{\rm Sh1}(y\tS{t+1}|x\tS{t+1}; \pi\tSb{t}) = \Ss_{\rm Sh1}(y\tS{t+1}|x\tS{t+1}; \pi\tSb{0})$ (c.f. \autoref{eq:Sh1}) which implies that $\Delta \Ss_{\rm Sh1}(y\tS{t+1}|x\tS{t+1}; \pi\tSb{t})$ (c.f. Proposition \ref{Prop:dSh_BF}) in \autoref{eq:prop_CC2} is equal to 0.
Similarly, in the limit of $p_c \to 1$, we have $\DKL \Big[ \pi\tSb{t} || p_c \pi\tSb{0} + (1-p_c)\pi\tSb{t} \Big] = \DKL \Big[ \pi\tSb{t} || \pi\tSb{0} \Big]$.
Therefore, in the limit of $p_c \to 1$ and given \autoref{eq:prop_CC2}, we have $\Ss_{\rm CC2}(y\tS{t+1}|x\tS{t+1}; \pi\tSb{t}) = \Ss_{\rm Ba2}(y\tS{t+1}|x\tS{t+1}; \pi\tSb{t})$.
$\hfill \blacksquare$

\subsection{Theoretical results for the postdictive surprise}

\begin{mylemma}
\label{Lemma:Po_Sh}
(Relation between the postdictive surprise and the Shannon surprise)
In the generative model of Definition \ref{def:gen_model},
the postdictive surprise can be written as
\begin{linenomath*} \begin{equation}\
\begin{aligned}
&\Ss_{\rm Po1}(y\tS{t+1}|x\tS{t+1}; \pi\tSb{t}) \\
&=\EE_{P\big(.|x\tS{t+1}; \PP\tSb{t}_{ \Theta\tS{t+1}} \big)}
\Big[
\Ss_{\rm Sh2}
\Big(
y\tS{t+1}|x\tS{t+1};
\PP\tSb{t}_{\Theta\tS{t+1}|Y,x\tS{t+1}}
\Big)
\Big]\\
&- \Ss_{\rm Sh1}(y\tS{t+1}|x\tS{t+1}; \pi\tSb{t})
\end{aligned}
\end{equation} \end{linenomath*}
and
\begin{linenomath*} \begin{equation}\
\begin{aligned}
&\Ss_{\rm Po2}(y\tS{t+1}|x\tS{t+1}; \pi\tSb{t}) \\
&= \EE_{P\big(.|x\tS{t+1}; \pi\tSb{t} \big)}
\Big[
\Ss_{\rm Sh2}
\Big(
y\tS{t+1}|x\tS{t+1};
\PP\tSb{t}_{\Theta\tS{t+1}|Y,x\tS{t+1}}
\Big)
\Big]\\
&- \Ss_{\rm Sh1}(y\tS{t+1}|x\tS{t+1}; \pi\tSb{t})\\
&+ D_{\rm KL} \Big[ P\big(.|x\tS{t+1}; \pi\tSb{t} \big) || P\big(.|x\tS{t+1}; \PP\tSb{t}_{\Theta\tS{t+1}} \big) \Big],
\end{aligned}
\end{equation} \end{linenomath*}
where $\PP\tSb{t}_{\Theta\tS{t+1}|y,x\tS{t+1}} \coloneqq \PP\tSb{t}_{\Theta\tS{t+1}}\big( .| Y\tS{t+1}=y , x\tS{t+1}\big)$ is the belief at time $t+1$ if we observe $Y\tS{t+1}=y$ with the cue $x\tS{t+1}$.
\end{mylemma}
According to Lemma \ref{Lemma:Po_Sh}, the postdictive surprise is equal to the difference between  the expected (over all values of $Y\tS{t+1}$) Shannon surprise of $Y\tS{t+2} = y\tS{t+1}$ given $X\tS{t+2} = x\tS{t+1}$ and the Shannon surprise of $y\tS{t+1}$ given $x\tS{t+1}$.

\textit{Proof:}
We first prove the equality for $\Ss_{\rm Po1}$ for which we have (c.f. \autoref{eq:Po1})
\begin{linenomath*} \begin{equation}\
\begin{aligned}
\label{eq:Po1_proof}
\Ss_{\rm Po1}&(y\tS{t+1}|x\tS{t+1}; \pi\tSb{t}) \\
&=
D_{\rm KL} \Big[ P\big(.|x\tS{t+1}; \PP\tSb{t}_{ \Theta\tS{t+1}} \big) ||  P\big(.|x\tS{t+1}; \pi\tSb{t+1} \big)  \Big]\\
&=
\EE_{P\big(.|x\tS{t+1}; \PP\tSb{t}_{ \Theta\tS{t+1}} \big)} \Big[ \log \frac
{P\big(Y|x\tS{t+1}; \PP\tSb{t}_{ \Theta\tS{t+1}} \big)}
{P\big(Y|x\tS{t+1}; \pi\tSb{t+1} \big)} \Big],
\end{aligned}
\end{equation} \end{linenomath*}
where
\begin{linenomath*} \begin{equation}\ \begin{aligned}
P\big(y|x\tS{t+1}; \PP\tSb{t}_{ \Theta\tS{t+1}} \big) &=
\int P_{Y|X}(y|x\tS{t}; \theta) \PP\tSb{t}\big( \Theta\tS{t+1} =\theta \big) d\theta,
\end{aligned} \end{equation} \end{linenomath*}
and, using Bayes' rule,
\begin{linenomath*} \begin{equation}\ \begin{aligned}
&P\big(y|x\tS{t+1}; \pi\tSb{t+1} \big) =
\int P_{Y|X}(y|x\tS{t+1}; \theta) \pi\tSb{t+1}(\theta) d\theta \\
&= \int P_{Y|X}(y|x\tS{t+1}; \theta)
\frac{\PP\tSb{t}\big( \Theta\tS{t+1} =\theta \big) P_{Y|X}(y\tS{t+1}|x\tS{t+1}; \theta)}{P\big(y\tS{t+1}|x\tS{t+1}; \PP\tSb{t}_{ \Theta\tS{t+1}} \big)} d\theta.
\end{aligned} \end{equation} \end{linenomath*}
Using the Bayes' rule and the definition of the marginal probability (c.f. \autoref{eq:marginal_prob}), we can find
\begin{linenomath*} \begin{equation}\ \begin{aligned}
&\frac
{P\big(y|x\tS{t+1}; \pi\tSb{t+1} \big)}
{P\big(y|x\tS{t+1}; \PP\tSb{t}_{ \Theta\tS{t+1}} \big)}
=
\frac{1}{P\big(y\tS{t+1}|x\tS{t+1}; \PP\tSb{t}_{ \Theta\tS{t+1}} \big)}\\
&\times \int P_{Y|X}(y\tS{t+1}|x\tS{t+1}; \theta)
\frac{\PP\tSb{t}\big( \Theta\tS{t+1} =\theta \big) P_{Y|X}(y|x\tS{t+1}; \theta)}{P\big(y|x\tS{t+1}; \PP\tSb{t}_{ \Theta\tS{t+1}} \big)} d\theta
\end{aligned} \end{equation} \end{linenomath*}
that is equal to
\begin{linenomath*} \begin{equation}\ \begin{aligned}
&\frac{\int P_{Y|X}(y\tS{t+1}|x\tS{t+1}; \theta) \PP\tSb{t}\big( \Theta\tS{t+1} =\theta | Y\tS{t+1}=y , x\tS{t+1}\big) d\theta}{P\big(y\tS{t+1}|x\tS{t+1}; \PP\tSb{t}_{ \Theta\tS{t+1}} \big)}\\
&\frac{\int P_{Y|X}(y\tS{t+1}|x\tS{t+1}; \theta) \PP\tSb{t}_{\Theta\tS{t+1}|y,x\tS{t+1}}(\theta) d\theta}{P\big(y\tS{t+1}|x\tS{t+1}; \PP\tSb{t}_{ \Theta\tS{t+1}} \big)}\\
=&
\frac{P\Big(y\tS{t+1}|x\tS{t+1}; \PP\tSb{t}_{\Theta\tS{t+1}|y,x\tS{t+1}} \Big)}
{P\big(y\tS{t+1}|x\tS{t+1}; \PP\tSb{t}_{ \Theta\tS{t+1}} \big)},
\end{aligned} \end{equation} \end{linenomath*}
and as a result (using \autoref{eq:Sh1} and \autoref{eq:Sh2})
\begin{linenomath*} \begin{equation}\ \begin{aligned}
\label{eq:proof_postdic_ratio}
\log &\frac
{P\big(y|x\tS{t+1}; \PP\tSb{t}_{ \Theta\tS{t+1}} \big)}
{P\big(y|x\tS{t+1}; \pi\tSb{t+1} \big)}
\\
=&- \log P\Big(y\tS{t+1}|x\tS{t+1}; \PP\tSb{t}_{\Theta\tS{t+1}|y,x\tS{t+1}} \Big)\\
&+ \log P\big(y\tS{t+1}|x\tS{t+1}; \PP\tSb{t}_{ \Theta\tS{t+1}} \big)\\
=&
\Ss_{\rm Sh2}(y\tS{t+1}|x\tS{t+1}; \PP\tSb{t}_{\Theta\tS{t+1}|y,x\tS{t+1}}   \big) \\
&- \Ss_{\rm Sh1}(y\tS{t+1}|x\tS{t+1}; \pi\tSb{t}),
\end{aligned} \end{equation} \end{linenomath*}
which, using \autoref{eq:Po1_proof}, makes the proof complete.

To prove the 2nd equality, we note that (c.f. \autoref{eq:Po2})
\begin{linenomath*} \begin{equation}\
\begin{aligned}
\label{eq:Po2_proof}
\Ss_{\rm Po2}&(y\tS{t+1}|x\tS{t+1}; \pi\tSb{t}) \\
&=D_{\rm KL} \Big[ P\big(.|x\tS{t+1}; \pi\tSb{t}  \big) ||  P\big(.|x\tS{t+1}; \pi\tSb{t+1} \big)  \Big]\\
&=
\EE_{P\big(.|x\tS{t+1}; \pi\tSb{t} \big)} \Big[ \log \frac
{P\big(Y|x\tS{t+1}; \pi\tSb{t} \big)}
{P\big(Y|x\tS{t+1}; \pi\tSb{t+1} \big)} \Big],
\end{aligned}
\end{equation} \end{linenomath*}
and
\begin{linenomath*} \begin{equation}\ \begin{aligned}
&\log \frac
{P\big(y|x\tS{t+1}; \pi\tSb{t} \big)}
{P\big(y|x\tS{t+1}; \pi\tSb{t+1} \big)}
=\\
&\log \frac
{P\big(y|x\tS{t+1}; \PP\tSb{t}_{ \Theta\tS{t+1}} \big)}
{P\big(y|x\tS{t+1}; \pi\tSb{t+1} \big)} +
\log \frac
{P\big(y|x\tS{t+1}; \pi\tSb{t} \big)}
{P\big(y|x\tS{t+1}; \PP\tSb{t}_{ \Theta\tS{t+1}} \big)}.
\end{aligned} \end{equation} \end{linenomath*}
Therefore, using \autoref{eq:proof_postdic_ratio} and the definition of $\DKL$, the proof is complete.
$\hfill \blacksquare$

\end{document}